\documentclass[twocolumn,trackchanges]{aastex62}

\usepackage{graphicx}
\usepackage{xcolor}
\usepackage[caption=false]{subfig}
\usepackage{booktabs}
\usepackage{txfonts}
\usepackage{textgreek}
\usepackage{multirow}
\usepackage{comment}

\graphicspath{{./}{figures/}}

\received{May 7, 2020}
\revised{September 15, 2020}
\accepted{September 24, 2020}
\submitjournal{ApJ}

\shorttitle{Earth as an Exoplanet: I. Time variable thermal emission using spatially resolved MODIS data}
\shortauthors{Mettler, Quanz \& Helled}

\begin{document}

\title{Earth as an Exoplanet: I. Time variable thermal emission using spatially resolved MODIS data}

\correspondingauthor{Jean-Noël Mettler}
\email{jmettler@phys.ethz.ch}

\author[0000-0002-8653-0226]{Jean-Noël Mettler}
\affiliation{ETH Zurich, Institute for Particle Physics and Astrophysics, Wolfgang-Pauli-Strasse 27, CH-8093 Zurich, Switzerland}
\affiliation{Center for Theoretical Astrophysics $\&$ Cosmology, Institute for Computational Science, University of Zurich, Switzerland}

\author[0000-0003-3829-7412]{Sascha P. Quanz}
\affiliation{ETH Zurich, Institute for Particle Physics and Astrophysics, Wolfgang-Pauli-Strasse 27, CH-8093 Zurich, Switzerland}

\author[0000-0001-5555-2652]{Ravit Helled}
\affiliation{Center for Theoretical Astrophysics $\&$ Cosmology, Institute for Computational Science, University of Zurich, Switzerland}

\begin{abstract}

Among the more than 4000 exoplanets known today, some terrestrial planets have  been detected in the so-called habitable zone of their host stars and their number is expected to increase in the near future, energizing a drive to understand and interpret the eagerly awaited wealth of data to identify signs of life beyond our Solar System. So far, Earth remains the best and only example of a habitable (and inhabited) world. Although, it seems extremely unlikely that any other exoplanets will be true Earth-twins, it is important to explore and understand the full range of spectral signatures and variability of Earth in order to inform the design of future instruments and missions, and  understand their diagnostic power as well as potential limitations.
In this work we use Earth observation data collected by the MODIS instrument aboard the Aqua satellite. The complete data set comprises 15 years of thermal emission observations in the 3.66--14.40~\textmu m range for five different locations on Earth (Amazon Rainforest, Antarctica, Arctic, Indian Ocean and the Sahara Desert). We then determine flux levels and variations as a function of wavelength and surface type (i.e. climate zone and surface thermal properties) and investigate whether periodic signals indicating Earth's tilted rotation axis can be detected. Our findings suggest that (1) viewing geometry plays an important role when thermal emission data is analyzed as Earth's spectrum varies by a factor of three and more depending on the dominant surface type underneath; (2) typically strong absorption bands from CO$_2$ (15 \textmu m) and O$_3$ (9.65 \textmu m) are significantly less pronounced and partially absent in data from the polar regions implying that estimating correct abundance levels for these molecules might be challenging in these cases; (3) the time-resolved thermal emission spectrum encodes information about seasons/planetary obliquity, but the significance depends on the viewing geometry and spectral band considered.

\end{abstract}


\section{Introduction}
\label{sec:intro}
Since the first detection of an exoplanet around a Sun-like star in 1995 \citep{Mayor1995}, the research field of exoplanets has grown rapidly. As of today, 25 years after the first discovery, there are over  4000 confirmed exoplanets in more than 3000 planetary systems \citep{alei2020}. Among these discoveries, some terrestrial planets have already been found orbiting in the so-called habitable zone (HZ) of their host stars such as planets e, f, and g in the TRAPPIST-1 system, LHS 1140~b, Proxima Centauri b, Kepler-442b or TOI-700d \citep{Gillon2017,Dittmann2017,Montet2015,Anglada2016,Gilbert2020}. In the future, their number is expected to increase further, primarily thanks to dedicated missions such as TESS \citep{TESS_2015} and PLATO \citep{Rauer_2014} and ongoing or upcoming radial velocity surveys such as CARMENES \citep{Quirrenbach_2014} or EXPRES \citep{Jurgenson_2016} and HARPS3 \citep{Young_2018}. 

The long-run goal is the characterization of temperate terrestrial exoplanets' atmospheres in order to assess their habitability and search for indications of biological activity. With the aim of detecting Earth-analogs around other stars, a particularly interesting class of exoplanets will be terrestrial planets  with thin N$_{2}$-H$_{2}$O-CO$_{2}$ atmospheres that orbit within the habitable zone of their host stars. Characterizing these planets and searching for traces of life requires the direct detection of their signal. The first space telescope capable of detecting potential atmospheres of terrestrial exoplanets is expected to be the James Webb Space Telescope (JWST) \citep[e.g.,][]{morley2017,totton2018,Tremblay2020}, which is scheduled for launch in 2021. From the ground, the 30-40m Extremely Large Telescopes may provide us with first images and  high-resolution spectroscopy data of a few terrestrial exoplanets, once they will be online in the mid to late 2020s \citep[e.g.,][]{quanz2015,snellen2015}. However, none of the currently planned missions or projects is capable of detecting and characterizing the atmospheres of a statistically meaningful sample of temperate, rocky exoplanets, and the exoplanet community has to wait for future missions like HabEx \citep{Gaudi2020}, LUVOIR \citep{LUVOIR_2019} or LIFE \citep{Quanz2018}.

For the time being, Earth remains the best -- and only -- example of a truly habitable and inhabited world, offering a unique opportunity to study the remote characterization of potential habitability. Despite the vast diversity of exoplanets \citep[e.g.,][]{Batalha2014, Burke2015} and the significant changes in the spectral appearance of Earth over the past 4.5 Gyr \citep[eg.,][]{Arney2016, Kasting_2003, Kaltenegger_2007, Meadows_2008, Reinhard_2017, Rugheimer_2018}, it seems  unlikely that any other exoplanet will be a true Earth-twin. Nevertheless, it is still useful to explore and understand the full range of spectral signatures and variability of Earth (e.g., due to the presence of oceans, clouds, surface inhomogeneities or polar caps) in order to fine-tune the design of future instruments and missions and understand their diagnostic power as well as potential limitations.

A key challenge arises from the fact that we will observe exoplanets from distances of -- typically -- at least several parsecs so that even with the most powerful telescopes conceived today they will remain spatially unresolved point sources. So, ideally, in order to study Earth as an exoplanet 'disk-integrated' data should be used. 
However, such data taken from a remote vantage point and showing the entire disk are limited as they are only obtained from a handful of spacecraft flybys such as \textit{Galileo}, \textit{MGS/TES}, \textit{EPOXI}, \textit{LCROSS} and \textit{DSCOVR} \citep[e.g.,][]{Sagan1993, Christensen1997, Livengood2011, Robinson2014, Yang2018}, spatially resolved satellite observations stitched together to a disk-integrated view of Earth \citep[e.g.,][]{Hearty2009} or from Earth-shine observations, where the reflected light from Earth is re-reflected from the unlit part of the Moon \citep[e.g.,][]{Turnbull2006, Palle_2009}. Nevertheless, the regular monitoring of Earth, in particular from Earth-orbiting satellites and to some extent also from other spacecraft, yielded a rich collection of Earth observation data, which has been used to investigate Earth's appearance in both reflected light and thermal emission \citep[for a recent review see, e.g.,][]{Robinson2018}. Such data offer an extensive temporal, spatial, and spectral coverage which can be tailored to represent exoplanet observation data or at least be interpreted in such a context. 

In this work, we use data collected by the Moderate Imaging Spectroradiometer (MODIS) aboard the Aqua satellite which has been orbiting Earth for the last 18 years. We focus on Earth's thermal emission of specific and representative surface types at mid-infrared (MIR) wavelength. By constructing data sets with a long time baseline spanning more than a decade and hence several orbital periods, we can investigate flux levels and variations as a function of wavelength range and surface type (i.e. climate zone and surface thermal properties) and look for periodic signals. 
Rotational and seasonal variations of Earth's spectrum and their influence on the detectability of spectral signatures of habitability and biosignatures have been investigated before \citep[e.g.,][]{Ford2001, Cowan2009,Fujii2011,Gomez2012,Hearty2009,Livengood2011,Robinson2011,Robinson2014,Jiang2018}, both based on observational data and simulations. Some studies also investigated Earth's thermal emission as an exoplanet using global climate models to simulate slowly rotating, ocean covered and snowball Earths or to estimate the eccentricity, obliquity and diurnal forcing of Earth-analogs \citep[e.g.,][]{Gomez2016, Cowan2012}.

While most studies so far focused on the reflected light spectrum, we here concentrate on the thermal emission of Earth and leverage long time baseline of the available data. Observing in the MIR enables us to characterize the thermal structure of exoplanetary atmospheres and provides additional information on the molecular abundances \cite[cf.][]{Hearty2009}. It also offers advantages over observing the reflected light of a planet with regard to seasonality. In this case, Mid-infrared will not be as negatively impacted by the lack of illumination of the winter hemisphere over the course of the orbit \cite[e.g., see Figure 5 in][]{Olson2018}. Ideally, for a comprehensive understanding of exoplanetary environments, combining the information from thermal emission and reflected light spectroscopy is needed.

We present spectral energy distributions (SEDs) and power spectral densities (PSDs) that were derived from Earth observation data for five specific locations on Earth. The data sets are constructed from 16 discrete channels located in the MIR (3.66--14.40~\textmu m) wavelength regime. To our knowledge, this is the longest continuously recorded time baseline from one satellite constellation using the same instrument for investigating Earth’s thermal emission in the context of exoplanet science.

\begin{figure*}[htb!]
    \centering
    \includegraphics[scale=0.33]{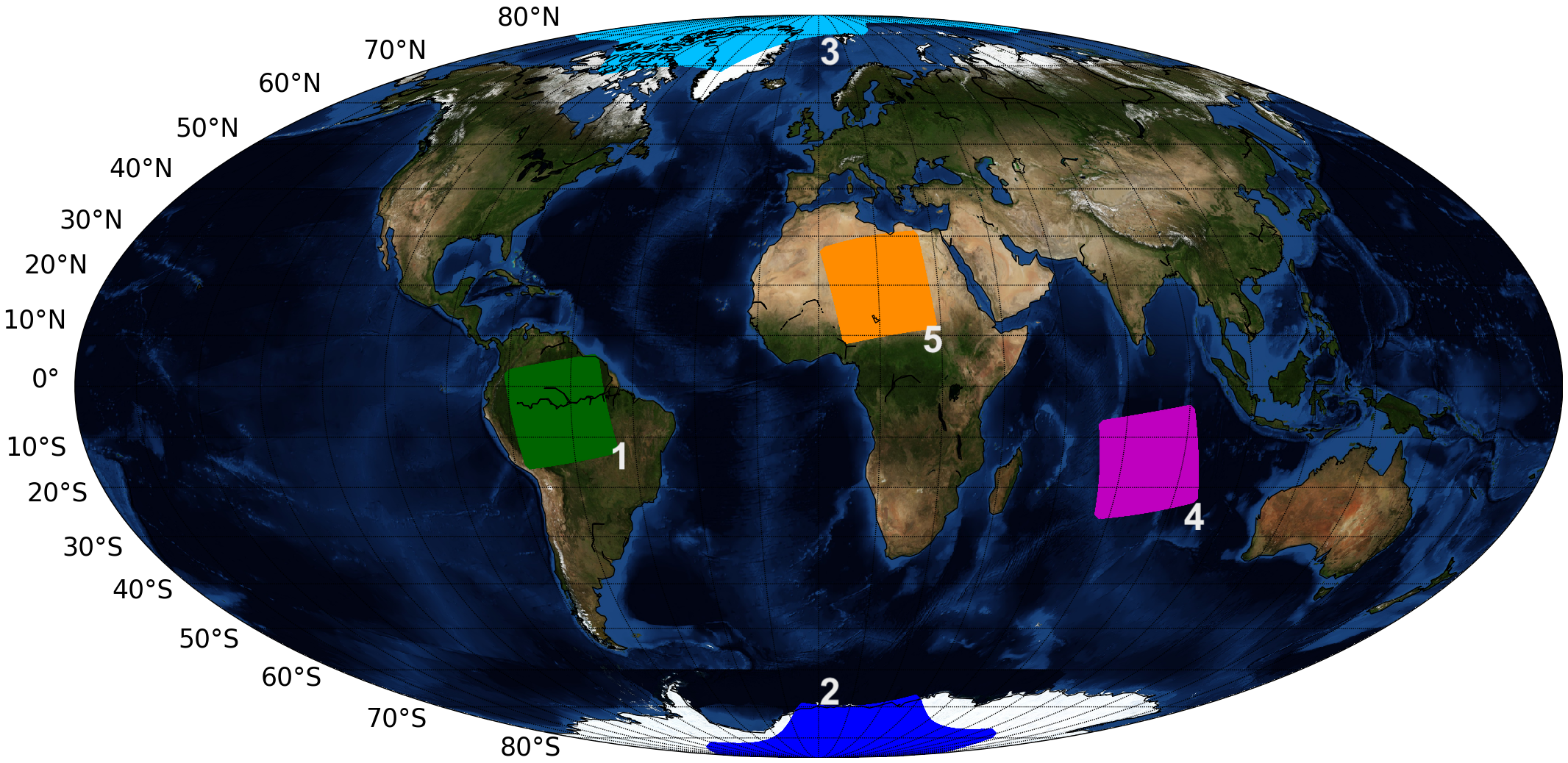}    
    \caption{World map highlighting the investigated target locations: (1) Amazon Rainforest, (2) Antarctica, (3) Arctic, (4) Indian Ocean and (5) Sahara Desert. At the equator, MODIS achieves a maximum swath size of 2330km by 2030km.}
    \label{Pic: Target Locations}
\end{figure*}

\newpage
\section{Methods}
\subsection{Locations and surface types} 

In order to investigate the thermal emission and time variability of Earth, we have focused on specific surface types with different thermal properties and from different climate zones that represent our planet’s appearance from afar. Although Earth shows a wide variety of surface characteristics, on  large scales it is dominated by oceans, deserts, ice and vegetation. In addition, clouds are also typically covering $\sim$ 67\% \citep{King2013} of the Earth. A comprehensive review about clouds and their effects on Earth's climate can be found in \citep{Kondratyev1999} and references therein.
In this paper we examine the following five locations: (1) Amazon Rainforest, (2) Antarctica, (3) Arctic, (4) Indian Ocean and (5) Sahara Desert as shown in Figure \ref{Pic: Target Locations}. For every location we also investigate the difference in thermal emission between day and night. 

The patch for the Sahara Desert covers the area between 10-30$^{\circ}$ N latitude and 0-30$^{\circ}$ E longitude. The time when the satellite was above this region varied over the 15 year period between 12:05 - 12:45 UTC and 01:00 - 02:00 UTC for the day and night measurements, respectively. The patches for the Amazon Rainforest were taken between 17:25 UTC - 18:05 UTC and 05:10 UTC - 05:40 UTC. For the Indian Ocean, we analyzed a patch lying below the equator between -10 and -25$^{\circ}$ S latitude and 65-90$^{\circ}$ E longitude. Here the time varied between 08:15 - 08:40 UTC for the day and 20:00 - 20:20 UTC for the night measurement. For the Arctic and Antarctica, the satellite was above the target region between 06:00 - 10:35 UTC and 16:30 - 18:55 UTC, respectively. Due to Earth's axial tilt the two polar data sets naturally include day and night measurements.


To ensure single surface type measurements consistent over the full time baseline, some patches had to be cut as described in section \ref{Sec: Data Retrieval}. Since we are interested in how Earth's appearance varies over time, we do not specifically select or de-select any data frames that are dominated by clouds. Instead, the cloud-induced variability in Earth's thermal emission is implicitly included in the analyses.

In order to interpret our analyses in the context of exoplanet science we assume (1) that for an exoplanet the surface types investigated here are dominating the hemisphere that is observed, and (2), related to this, that the integration time in order to collect the exoplanet data is short compared to the rotation period of the planet so that "smearing effects", i.e., the mixing of various surface types, are negligible.

\subsection{Data retrieval and processing}
\label{Sec: Data Retrieval}
We used data collected by MODIS, a cross-track, multi-disciplinary scanning radiometer with in total 36 spectral channels measuring visible and infrared bands in the wavelength range of 0.4--14.4~\textmu m. Approximately 40 data products are produced from MODIS data with horizontal spatial resolutions at nadir of 250 m, 500 m and 1000 m, providing the most detailed spatial information of all instruments aboard the Aqua satellite \citep{Aqua_Brochure}. The instrument employs a conventional imaging spectroradiometer concept, consisting of a cross-track scan mirror and collecting optics, and a set of linear arrays with spectral interference filters located in four focal planes. The optical arrangement provides high radiometric sensitivity (12 bit) imagery throughout the channels. With a temporal resolution of 5 minutes, MODIS achieves a maximum swath of 2330 km by 2030 km at the equator due to a $\pm$ 55-degree scanning pattern. This leads to low-latitude data gaps between successive orbits, preventing full global daytime or nighttime coverage within a day. Although these data gaps are filled in on subsequent days, the full global coverage is obtainable every two days. However, due to Aqua's operation height of 705 km and due to its sun-synchronous, near-polar and circular orbit, it revolves Earth in 99 minutes with a repeat cycle of 16 days. 
Therefore, our target locations were monitored twice per month, resulting in 360 data points per location after 15 years of observation.

\par

We investigated the calibrated at-aperture radiance for 16 thermal channels (see, Table~\ref{Table: Thermal Bands}) with a spatial resolution of 1000 m that were measured twice per month between January 2003 and December 2017. The data originate from a Level-1B product called \textit{MODIS / Aqua Calibrated Radiances 5-Min L1B Swath 1km} (MYD021KM) collection number 6.1 \href{http://dx.doi.org/10.5067/MODIS/MYD021KM.061}{(LAADS DAAC)}, which was accessed over NASA's  \href{https://search.earthdata.nasa.gov}{\textit{EARTHDATA Engine}}. This product level results from applying sensor calibration to the raw radiance counts (Level-1A) in order to produce calibrated, top-of-atmosphere (TOA) radiances.

\begin{table}[htb!]
\caption{Summary of all thermal channels of MODIS. Channel 26 covers the 1.360--1.390~\textmu m range and does not probe the Earth's thermal emission. Therefore, it is not listed below. Taken from \href{https://modis.gsfc.nasa.gov/about/design.php}{https://modis.gsfc.nasa.gov/about/design.php}.\label{Table: Thermal Bands}}
\small
\centering          
\begin{tabular}{l c c}
\toprule
\toprule
\textbf{Primary Use} & \textbf{Channel} & \textbf{Bandwidth [\textmu m]} \\ 
\midrule
\multirow{4}{2.5cm}{Surface / Cloud Temperature} &	20 & 3.660 - 3.840\\
 & 21 & 3.929 - 3.989\\ 

 & 22 & 3.929 - 3.989\\ 
 
 & 23 & 4.020 - 4.080\\ \midrule

\multirow{2}{2.5cm}{Atmospheric Temperature} & 24 & 4.433 - 4.498\\ 
 
 & 25 & 4.482 - 4.549\\ \midrule


\multirow{2}{2.5cm}{Cirrus Clouds / Water Vapour}& 27 & 6.535 - 6.895\\ 

 & 28 & 7.175 - 7.475 \\ \midrule
 
Cloud Properties & 29 & 8.400 - 8.700 \\ \midrule

Ozone & 30 & 9.580 - 9.880 \\ \midrule

\multirow{2}{2.5cm}{Surface / Cloud Temperature} & 31 & 10.780 - 11.280 \\ 

 & 32 & 11.770 - 12.270 \\ \midrule
 
 \multirow{4}{2.5cm}{Cloud Top Altitude} & 33 & 13.185 - 13.485 \\ 

 & 34 & 13.485 - 13.785\\ 
 
 & 35 & 13.785 - 14.085\\ 
 
 & 36 & 14.085 - 14.385 \\ 
\bottomrule
\end{tabular}
\end{table}

After downloading the corresponding data from the servers, the files were read and the band specific scaled integers (\textit{SI$_{B}$}) were converted into spectral radiances ($L_{B}$) with physical units of W m$^{-2}$\textmu m$^{-1}$ sr$^{-1}$ according to equation \ref{Eqn: SI Conversion} \citep{MODIS_Userguide}
\begin{equation}
\centering
L_{B} = \Gamma_{B} (SI_{B} - \Theta_{B}),
\label{Eqn: SI Conversion}
\end{equation}
where \textit{$\Gamma_{B}$} and \textit{$\Theta_{B}$} refer to radiance scales and radiance offsets for a specific band \textit{B}, respectively, which are computed inside Level-1B products and are written as attributes to the scientific data sets (SDS). In the MYD021KM Level-1B product, the radiance measured in each channel were generated in 32-bit floating-point format and scaled to a range of [0, 32767]. Any value larger than 32767 represents invalid or unusable data and was removed from the data sets.

The first iteration through this process revealed that in some years the patches of the selected locations were slightly drifting. These drifts are caused by orbital maneuvers which are well known \citep[e.g.,][]{Lin2019}. In order to solve this issue we had to custom clip the concerning patches to show the exact same location for each passage. While only a few frames of the Sahara Desert had to be slightly cut, an effort was made to cut the two polar regions patches in a way that the geographical pole lies in the center. As a result, roughly 50\% of the Arctic patch had to be clipped to just show the polar cap, thus, instead of having the usual 1354 by 2030 pixels the scene is composed of 600 by 2030 pixels. None of the Amazon Rainforest or Indian Ocean patches had to be cut as they all showed scenes composed of vegetation and water (and clouds), respectively.


\section{Results}
\label{Sec: Results}

\subsection{Spectral energy distributions (SEDs)} 
\label{SubSec: SEDs}
\begin{figure}[htb!]
    \vspace{0.5cm}
    \centering
    \includegraphics[width=\hsize]{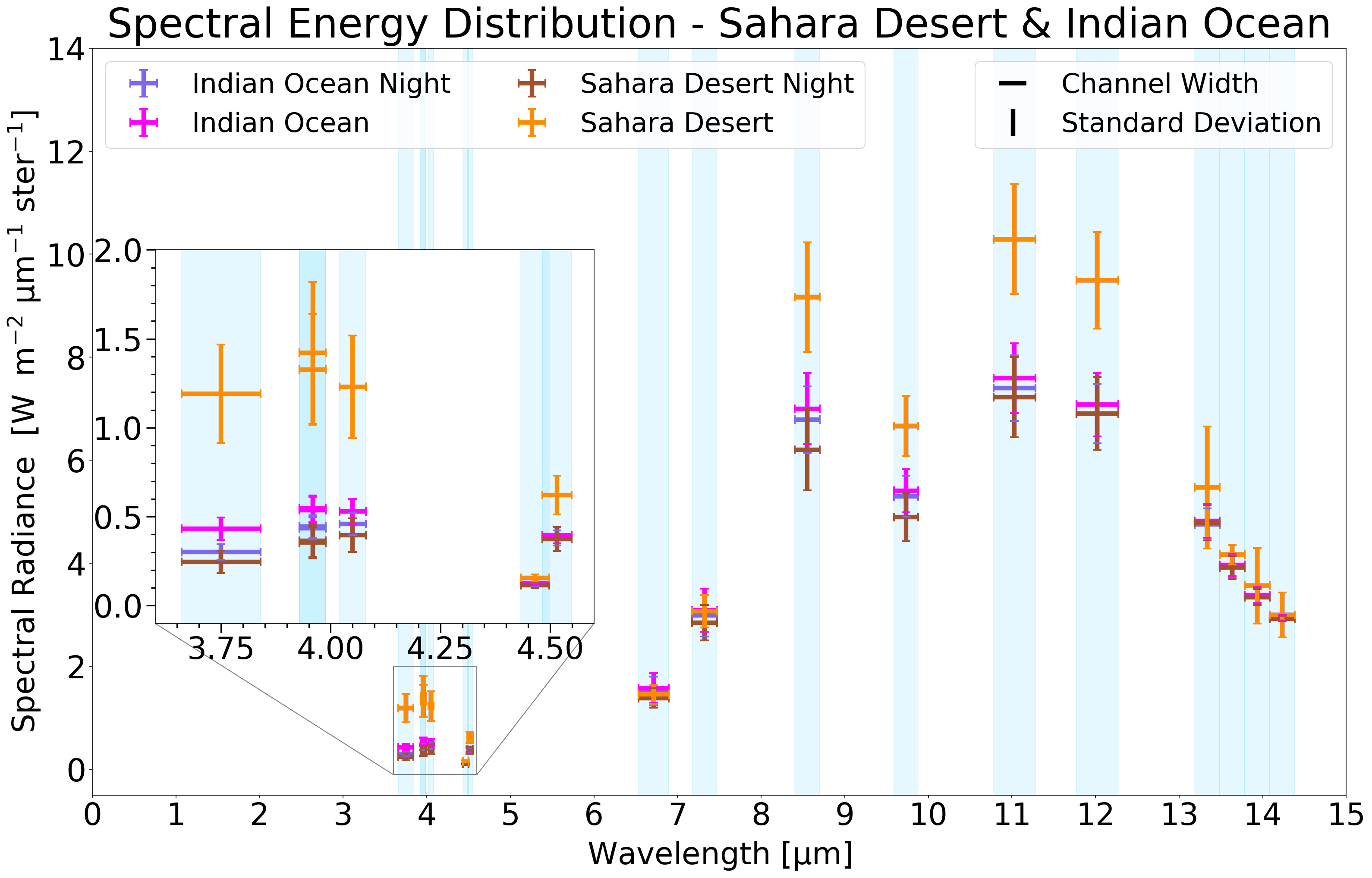}
    \caption{Spectral energy distributions of the Sahara Desert and the Indian Ocean. Their night measurements are shown in sienna and dark green, respectively. The horizontal lines in x-direction correspond to the average radiance measured in every channel over the 15-year observation period and the error bars in y-direction to the standard deviation. The blue shadings indicate the bandwidth of each channel.}

    \label{Pic: SEDs Equatorial}
\end{figure}

\begin{figure}[htb!]
\centering
    \includegraphics[width=\hsize]{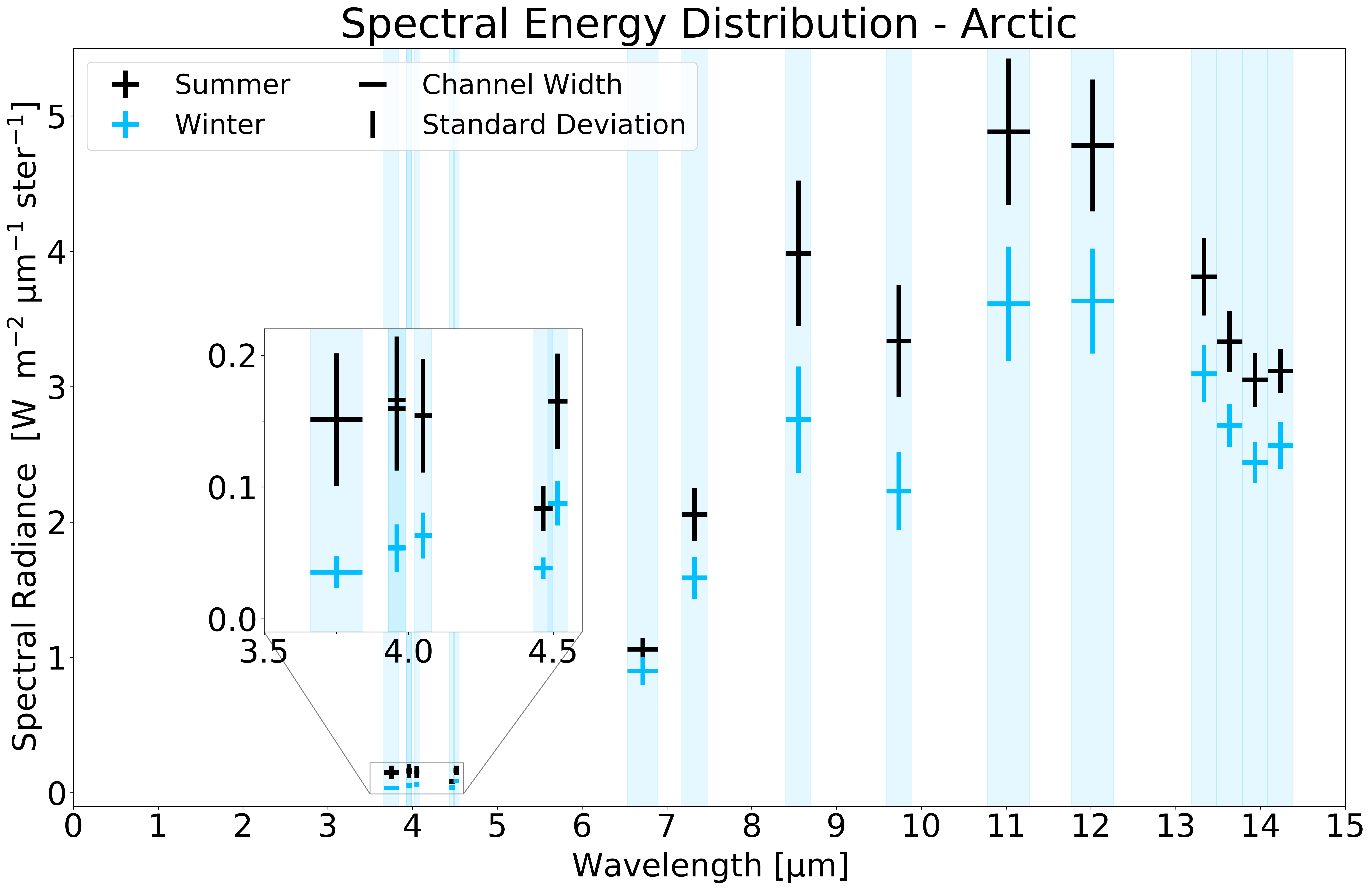}
    \caption{Spectral energy distribution of the Arctic. The summer data is shown in black and includes data from April - September and winter (sky blue) from October - March. The increase towards the end of the spectrum in channel 36 (14.085 - 14.385 \textmu m) is a characteristic feature for this type of location and indicates the emission of CO$_{2}$ centered at 15 \textmu m.}
    \label{Pic: SED Arctic}
\end{figure}

\begin{figure}[htb!]
\vspace{0.5cm}
\centering
    \includegraphics[width=\hsize]{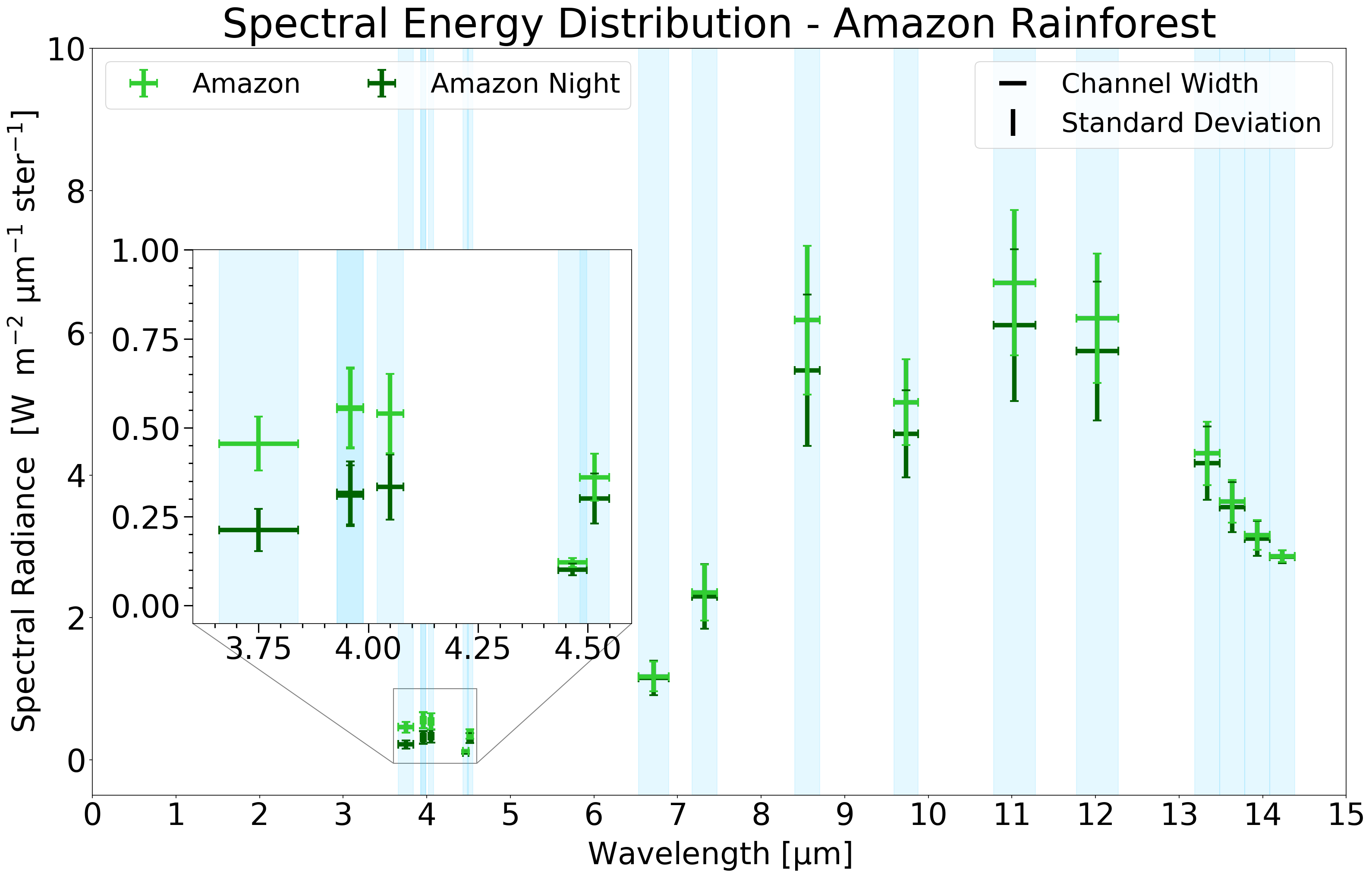}
    \caption{Spectral energy distribution of the Amazon Rainforest displaying the major absorption features of ozone at 9.65 \textmu m and carbon dioxide at 15 \textmu m. }
    \label{Pic: SED Amazon}
\vspace{1.3cm}
\end{figure}

\begin{figure}[htb!]
\vspace{0.5cm}
\centering
    \includegraphics[width=\hsize]{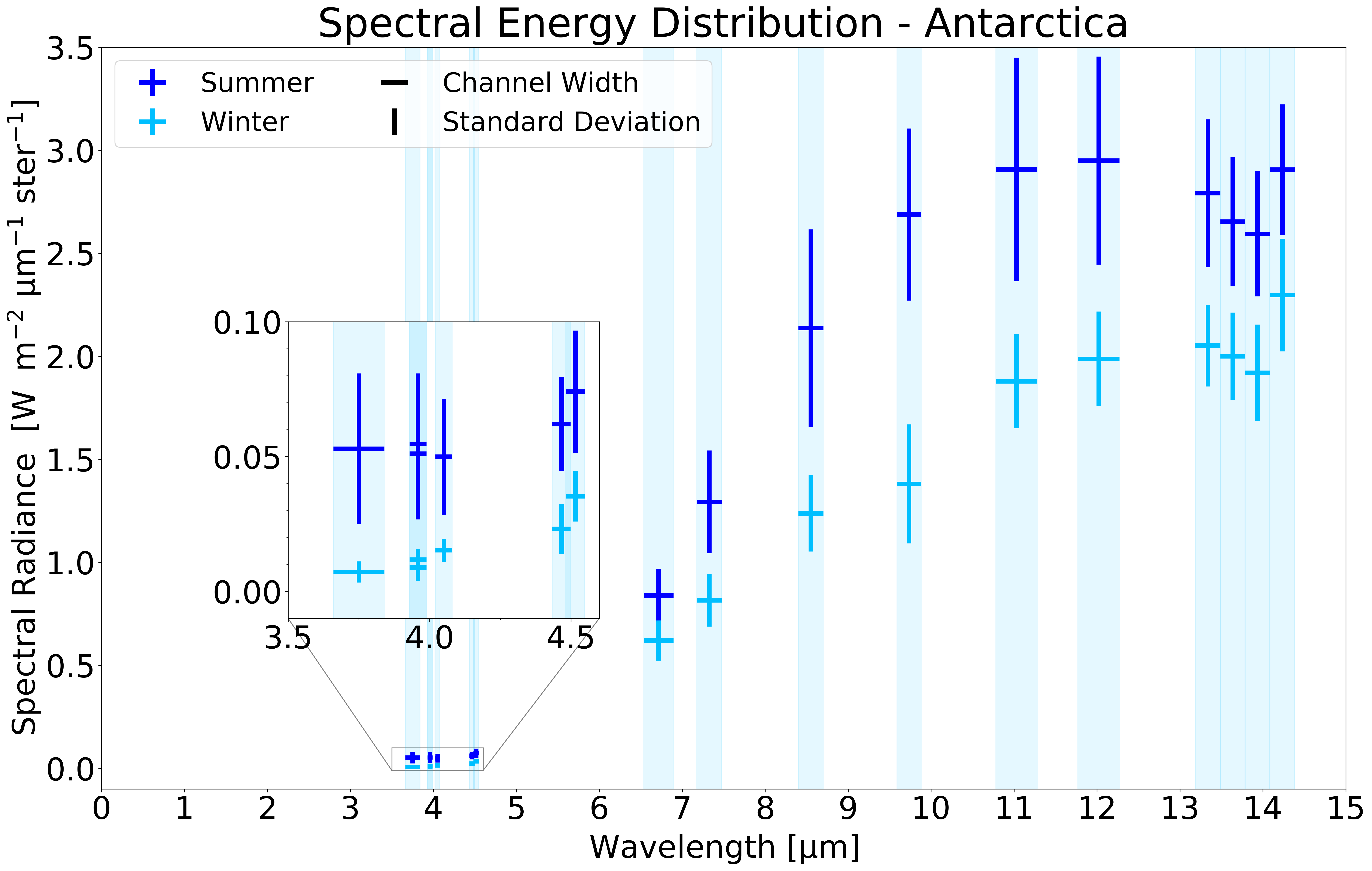}
    \caption{Spectral energy distribution of the Antarctica. Same as Figure \ref{Pic: SED Arctic} but the definition of summer (blue) and winter (sky blue) is switched. 
    Like the Arctic, the SED shows an upturn in flux in the last channel for both seasons.}
    \label{Pic: SED Antarctica}
\end{figure}

A key diagnostic for investigating the (atmospheric) properties of an exoplanet is its spectral energy distribution (SED). In Figures ~\ref{Pic: SEDs Equatorial} - \ref{Pic: SED Antarctica} we show the SEDs of all target locations including their day and night comparisons (note the different scales on the y-axes). In appendix \ref{Appendix} we provide the mean values for all SEDs in tabulated form. The horizontal lines indicate the average radiance measured in every channel over the 15-year observation period. The length of the lines in x-direction corresponds to the channel bandwidth and the error bars in y-direction to the measured standard deviation of the radiance. The latter encodes a mix of intrinsic variability for a given location at a certain time of the year and variations due to seasonal effects (see, Figure \ref{Pic: spec rad comp arctic antarc}). In order to derive the day and night SEDs for the polar regions, their data was split into summer and winter data with each season lasting for 6 months.


For the north pole, summer was defined as from April - September and winter from October - March and vice versa for the south pole. As expected, we find the hottest region, the Sahara Desert, with the highest radiance values followed by the Indian Ocean, Amazon Rainforest, Arctic, and, finally the coldest target location, Antarctica. Throughout the channels and over the entire observation period of 15 years, the Sahara Desert also shows the largest difference between day and night measurements where the flux differs by 30$\%$ in channel 31 (10.78 - 11.28 \textmu m). The Indian Ocean data lies in between this range and overall shows a smaller dispersion which could be due to the oceanic large thermal inertia (heat capacity) making them more resistant to temperature changes. 

We note the very low level of variations at the shortest wavelength for the two poles and the Indian Ocean which can largely be attributed to the overall low flux levels. In general, comparing the radiance from the equatorial regions to that from the polar regions, the flux differs by up to a factor of about two and three during the day and night, respectively. However, this strongly depends on the channel and location considered.

In order to estimate the effective temperature of each target location, we fitted a black-body curve to every single measurement's SED and calculated the weighted arithmetic mean and the corresponding standard error of the weighted mean from the total set. The results are shown in Table \ref{Table: Fitted Temps}. 
Although no comparable results were found that cover the same observation period, the obtained estimates compare well to daily measurements of the same regions \cite[e.g.,][]{Hearty2009,Hanel1972,Petitcolin2002}. 

\begin{table}[!htb]
\caption{Overview of the resulting effective temperatures after a black-body curve was fitted to every single measurement's SED of a target location and the weighted arithmetic mean and the standard error of the weighted mean were calculated from the set. For the Arctic, summer and winter were defined as from April - September and October - March, respectively, and vice versa for Antarctica.}
\label{Table: Fitted Temps}
\centering
\small
{\def\arraystretch{1.2}
\begin{tabular}{l c}
\toprule
\toprule
\textbf{Target Location} & \textbf{Effective Temperature [K]} \vspace{0.1cm}\\
\midrule
\vspace{0.1cm}
Sahara Desert& 278.01 $\pm$ 0.31\\ 
Indian Ocean & 267.92 $\pm$ 0.21 \vspace{0.1cm}\\
Amazon Rainforest & 260.52 $\pm$ 0.20 \vspace{0.1cm}\\
Arctic Summer& 249.18 $\pm$ 0.16 \vspace{0.1cm}\\ 
Antarctica Summer& 233.83 $\pm$ 0.06 \vspace{0.3cm}\\ 

Sahara Desert Night & 264.39 $\pm$ 0.20 \vspace{0.1cm}\\ 
Indian Ocean Night & 266.82 $\pm$ 0.20 \vspace{0.1cm}\\ 
Amazon Rainforest Night & 255.83 $\pm$ 0.17 \vspace{0.1cm}\\
Arctic Winter& 236.75 $\pm$ 0.12 \vspace{0.1cm}\\ 
Antarctica Winter& 219.00 $\pm$ 0.02 \vspace{0.1cm}\\ 
\bottomrule
\end{tabular} }
\end{table}

Four out of five locations, the Sahara Desert, Indian Ocean, Amazon Rainforest and the Arctic, show similar spectral absorption features in their SEDs. The first feature around  $\approx$ 4.475~\textmu m in channel 24 (4.433 - 4.498~\textmu m) corresponds to the CO$_{2}$ band centered at 4.3 \textmu m \citep{Catling2018}. It is followed by a clearly visible ozone absorption band at 9.65~\textmu m, and between channel 33 and 36 (13.2 - 14.4~\textmu m), for the three equatorial regions, the plots show a continues decrease which indicates the next absorption band due to CO$_{2}$ centered at 15~\textmu m \cite[e.g., Figure 3 in][]{Schwieterman2018, Catling2018}. 

The two polar regions behave rather different to each other as the south pole displays a seasonal dependence when the drop in variability in channel 30 (centered on the ozone band at 9.65~\textmu m) relative to channels 29 and 31 is compared. During the southern hemisphere summer, the resulting SED for Antarctica (shown in Figure \ref{Pic: SED Antarctica}) is missing the dominant ozone absorption feature at 9.65 \textmu m. An existing but less pronounced ozone feature is, however, observable during winter. Due to the missing ozone absorption feature in the summer data, the SED of Antarctica has a shape similar to that of a perfect black-body with a temperature of 233.83 K $\pm$ 0.06 K. Also, both polar regions show an upturn in flux towards the end of the observed wavelength range by MODIS.

\begin{figure}[htb!]
\centering
    \includegraphics[width=\hsize]{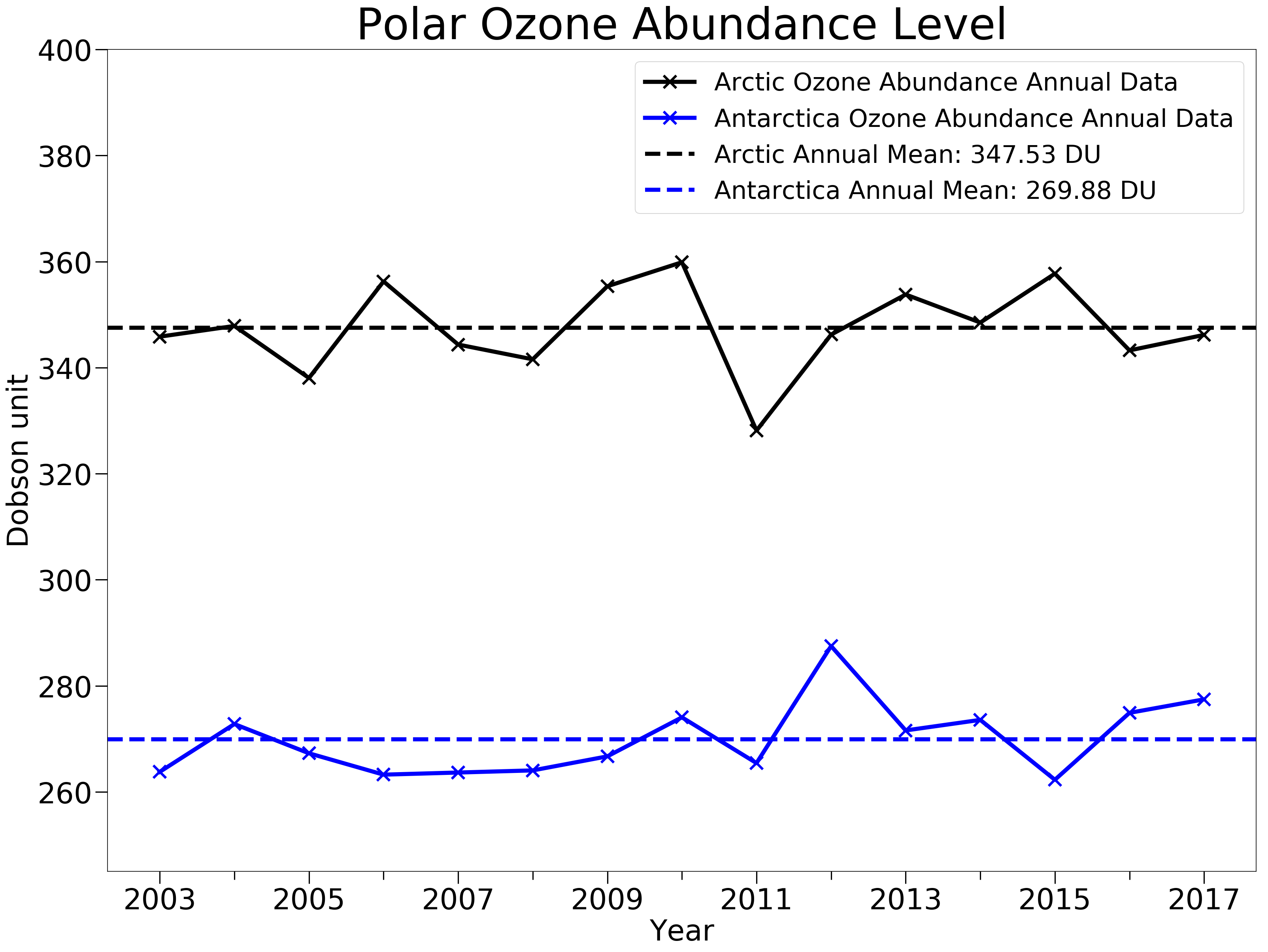}
    \caption{A comparison between the Arctic and Antarctic ozone abundance over the same 15-year period that was used to derive the SEDs. Within that time period, the data show that there is 22$\%$ less ozone in the south than in the north.}
    \label{Pic: Ozone Abundance}
\end{figure}

A reason for the difference in depth of the ozone absorption feature between the two poles could be the fact that ozone is less abundant in the southern hemisphere. Although, a thinning of the ozone layer has been observed over other regions \cite[e.g.,][and references therein]{Butchart2014}, such as the Arctic as well as northern and southern midlatitudes, the most severe ozone loss was registered to be recurring in springtime over Antarctica due to chlorofluorocarbons, known as CFCs, that lead to depletion of the ozone layer \citep{Molina1974}. Figure \ref{Pic: Ozone Abundance} shows an ozone abundance data set for both polar regions over the same 15-year time baseline used in our SED analysis. The data set was provided by NASA's Ozone Watch. The 15 year Arctic ozone abundance average is calculated to be 347.53 Dobson Units (DU) and 269.88 DU for the Antarctica, indicating that there is on average 22$\%$ less ozone around the south pole compared to the north pole within that time period.

 Despite the fact that such a shape, and thus temperature, would originate from high altitude clouds, this can be ruled out due to the long observation period. While less ozone is certainly present in the atmosphere above Antarctica, we primarily attribute the lack of a clear absorption feature to the temperature structure of the atmosphere at this location rather than due to differences in abundance. Balloon soundings showed that 90$\%$ of the atmospheric ozone resides in the stratosphere between 10 - 17 km and about 50 km with a peaking abundance around 20 - 25 km. At this altitude the average temperature above the south pole over a 40 year period is roughly $\approx$ 221~K (NASA Ozone Watch). With the surface temperature being similar to the temperature at the altitude of peak ozone abundance, the absorption feature may not be seen due to the missing contrast in thermal emission.

The observed upturn in channel 36 (14.085 - 14.385 \textmu m) in the SEDs for both polar regions appears to be a characteristic signature of these types of locations. For the poles, it has been known for years that surface-based temperature inversions exist \cite[e.g.,][]{Hudson2005} which are major drivers of some aspects of their climates, including atmospheric radiation. Given that the snow-surface emissivity is greater than the atmospheric emissivity, the inequality in emissivities results in a temperature inversion when the absorbed energy from solar radiation at the surface is small. In this case, some features can appear in emission. Hence, the increase in channel 36 indicates the emission of CO$_{2}$ centered at 15 \textmu m.

\subsection{Power Spectral Densities}
\label{SubSec: PSDs}

\begin{figure}[htb!]
    \centering
    \includegraphics[width=\columnwidth]{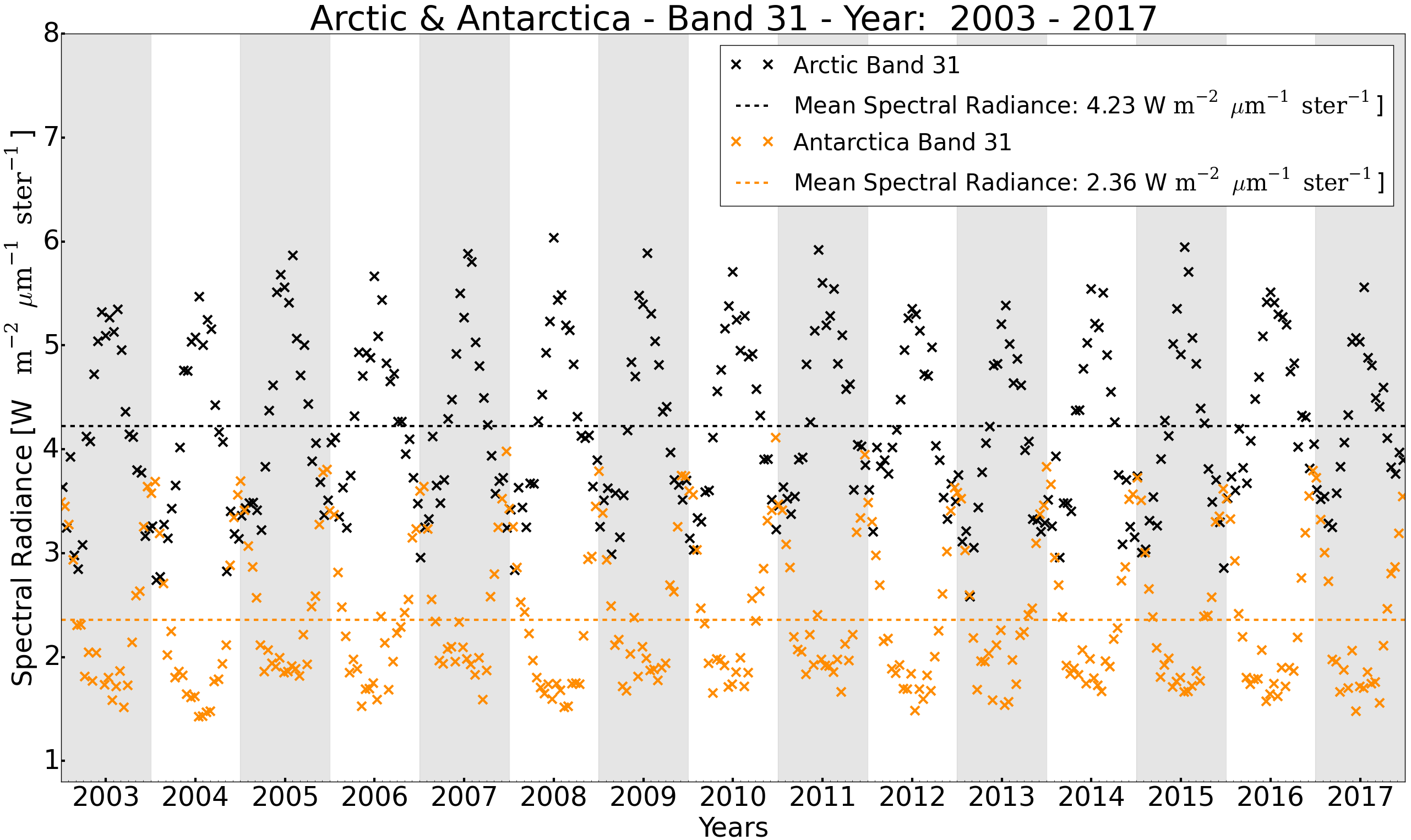}
    \caption{Comparison between the two polar regions (Arctic black data points, Antarctica orange data points) for channel 31. Due to Earth's obliquity, the poles do not receive the same insolation at the same time, leading to higher (lower) spectral radiance readings at the north (south) pole during the summer, as seen from the northern hemisphere. The mean radiance level over the 15 years period is indicated by the dashed horizontal lines. The mean spectral radiance difference between the two poles is 1.87 W m$^{-2}$ \textmu m$^{-1}$ sr$^{-1}$.}
    \label{Pic: spec rad comp arctic antarc}
\end{figure}

In order to investigate whether the planet's obliquity can be inferred from Earth's thermal emission, we decompose the spectral radiance time series into components of different frequencies using Fourier analysis. Specifically, we performed a Discrete Fourier Transform (DFT) on the data to produce a power spectrum, also known as power spectral density (PSD). Since the PSD is the measure of a signal's power content versus frequency, finding trends in the data related to the orbital period would be a strong indication for a tilted spin axis. Figure \ref{Pic: spec rad comp arctic antarc} gives an example of the variations in spectral radiance. It shows the time series of the observed spectral radiance in channel 31 (10.78 - 11.28~\textmu m) for the patches over the poles revealing a cyclic behaviour. Due to Earth's obliquity, the poles show higher (lower) spectral radiance readings at the north (south) pole during summer (winter). Fitting a sine function to the data revealed a periodicity of $\approx$365.25 days, which is very close to the official sidereal year with  $\approx$365.2564 SI days \citep{IERS}. A comparison on the average radiance of the different surface types shows that land masses emit more radiation than ocean or ice covered areas, which is in agreement with previous studies \cite[e.g.,][]{Hearty2009, Gomez2012, HURLEY2014}. 

\begin{figure*}[htb!]
	\centering
    \includegraphics[width=0.90\hsize]{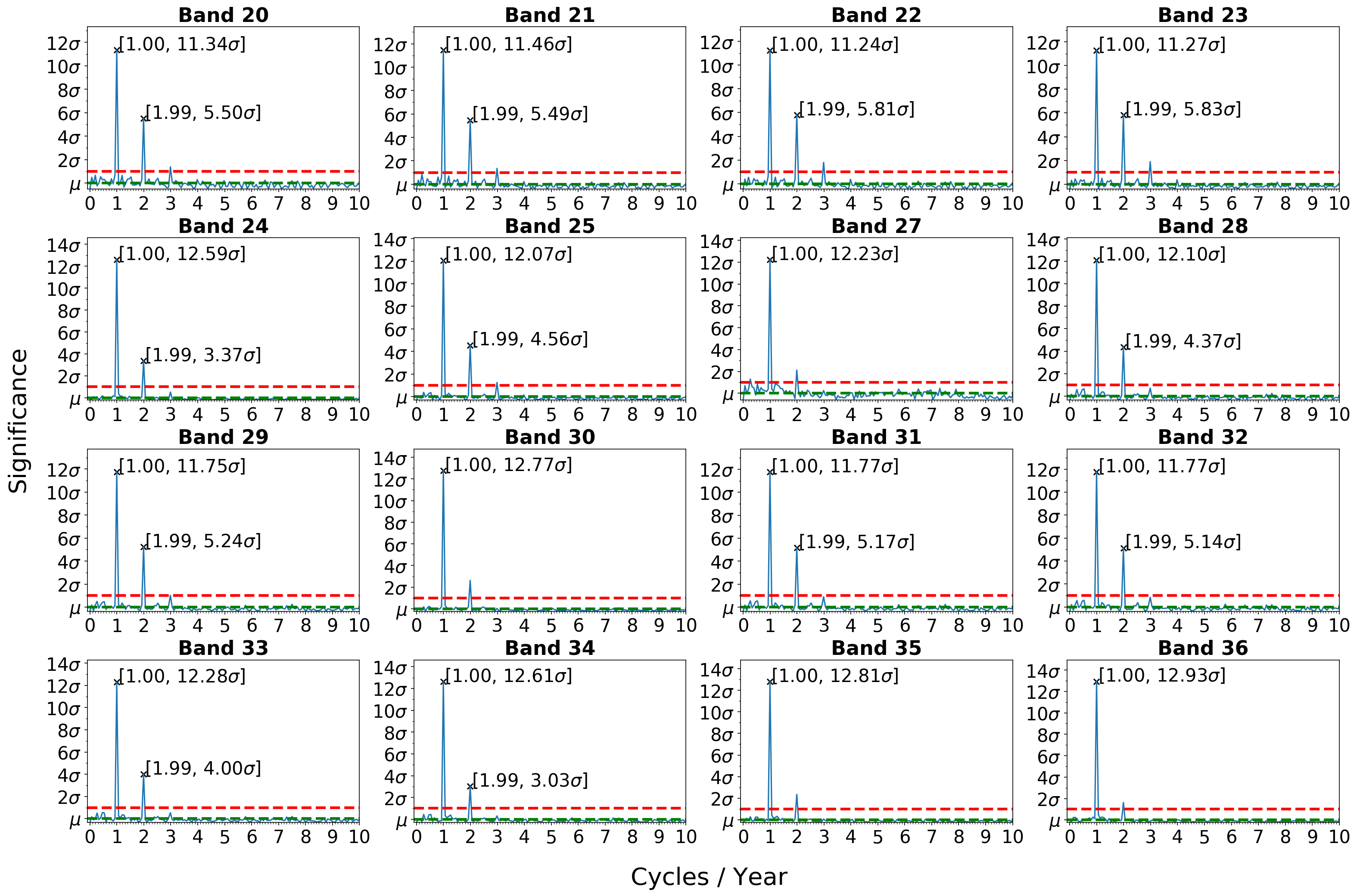}
    \caption{Calculated power spectral density for the Antarctica data set. The figure shows cycles / year on the x- and the significance ($\sigma$) of the signal on the y-axis. The green and red dashed lines correspond to the signal's mean and $\pm$ 1$\sigma$ value, respectively. Due to the clear periodicity in the spectral radiance signal caused by the Earth's obliquity, a relative clean PSD with a noise level well below 1$\sigma$ can be calculated for the polar regions. All channels show a strong peak (above 11.2$\sigma$) at 1 cycle / year and are therefore indicating a tilted rotation axis of Earth. Due to the well defined peaks, all channels are appropriate for looking for evidence of planetary obliquity in the MIR spectrum for the polar regions.}
    \label{Pic: PSD Antarctica in text}
\end{figure*}

A DFT was applied to every channel for each target location by using python's numpy \textit{numpy.fft.fft} module which calculates the one dimensional n-point DFT with the efficient Fast Fourier Transform (FFT) algorithm. As it is a common practice to express the significance of an enhancement by quoting the number of standard deviations it differs from the mean value of the signal, Figure \ref{Pic: PSD Antarctica in text} shows cycles per year on the x-axis and the significance of a possible signal on the y-axis. It should be noted that the following discussion is based on the results where a DFT is applied to a data set with a time baseline of 15 years. The results of a one year baseline are discussed further below. 

\begin{figure}
    \vspace{0.5cm}
	\centering
        \includegraphics[width=\columnwidth]{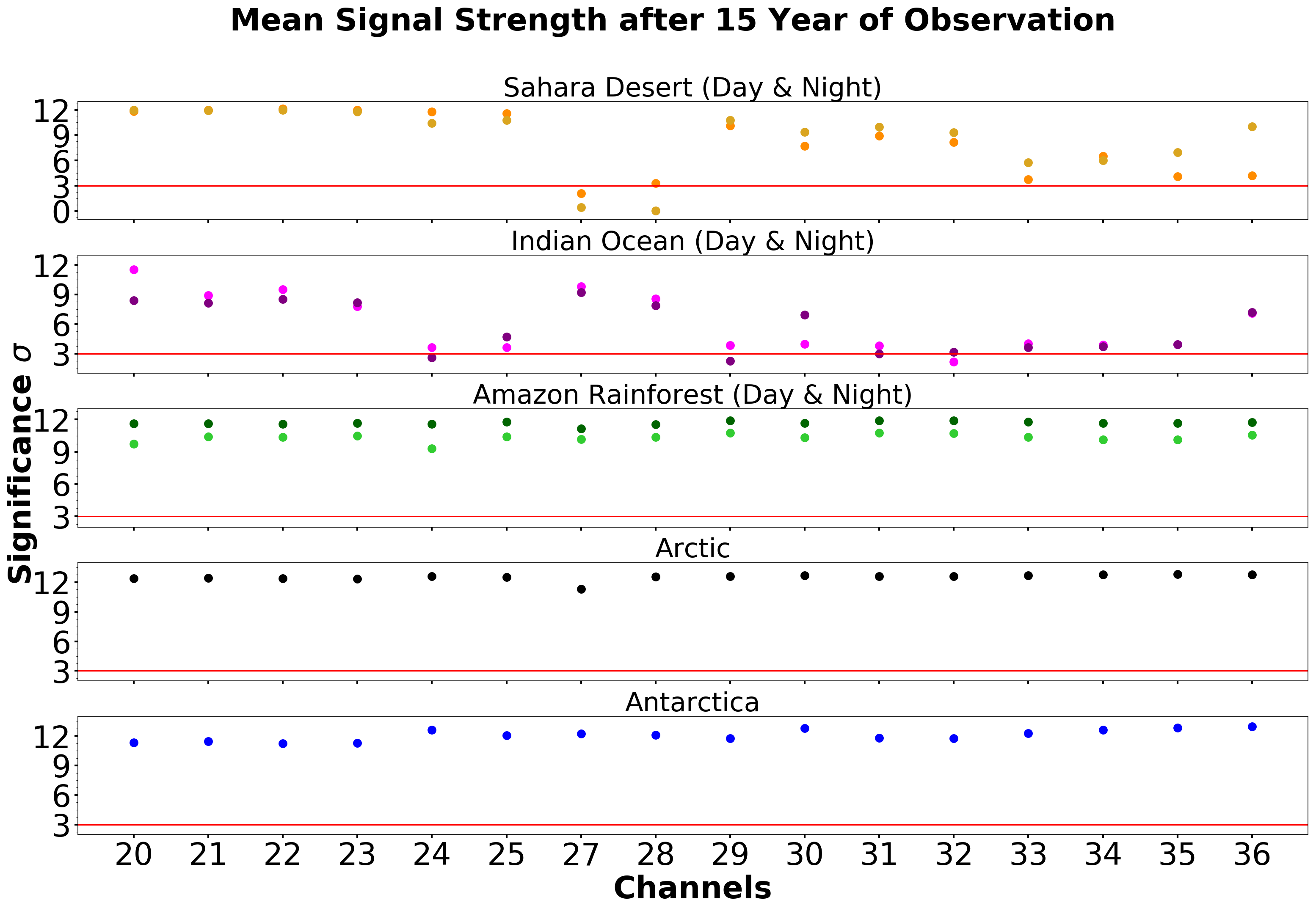}
        \caption{PSD analysis for 15 consecutive years of observation to infer the existence of seasons. The figure shows the 16 thermal channels on the x- and the significance $\sigma$ on the y-axis. The red line indicates the 3$\sigma$ threshold and the errorbars correspond to the channel specific standard deviation. All channels can be used to infer seasons for the polar regions and the Amazon Rainforest  as well as the majority of the channels for the Sahara Desert and Indian Ocean.}
        \label{Pic: PSD 15yr Observation}
\end{figure}

Although the power spectral densities are noisier for the equatorial regions compared to the polar regions throughout all available thermal channels, 92$\%$ of all channels and locations including night measurements show a well defined peak (3$\sigma$ and more) at 1 cycle per year as displayed in Figure \ref{Pic: PSD 15yr Observation}. The polar regions show overall peaks of 10$\sigma$ or more as well as whole integer repeats which can be interpreted as harmonics, thus artefacts of the method. For these two locations, all thermal MODIS channels seem to be appropriate for looking for evidence of planetary obliquity in the MIR spectrum. Even though some bands are location and surface type dependent, we find four channels (20-23, 03.66-04.08~\textmu m) that produce strong signals (7$\sigma$ and above) showing evidence of planetary obliquity independent of target location and surface type. These channels are primarily used for surface and cloud temperature measurements as stated in Table \ref{Table: Thermal Bands}. 

\begin{figure}
    \vspace{0.5cm}
    \includegraphics[width=\columnwidth]{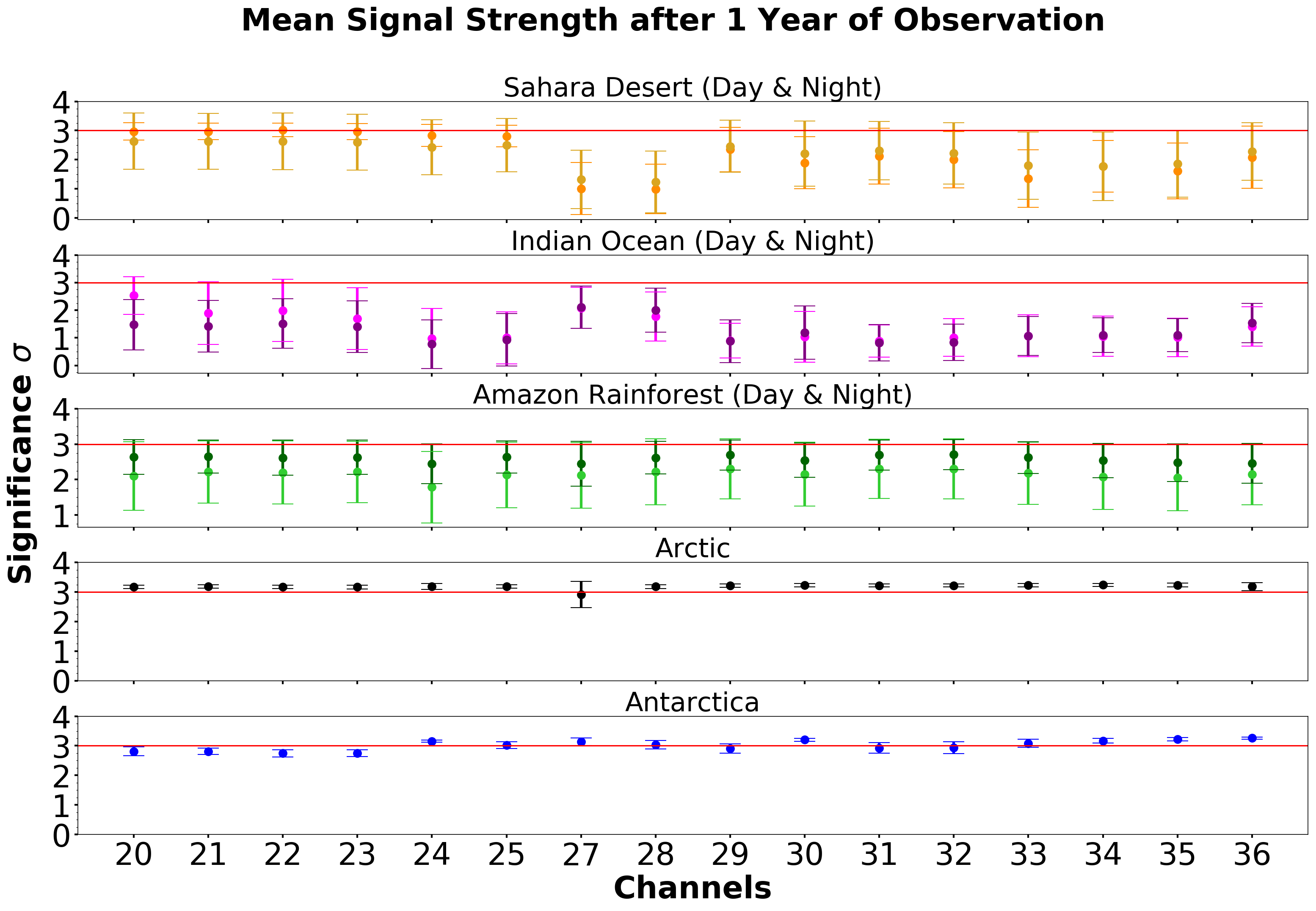}
    \caption{Same as Figure~\ref{Pic: PSD 15yr Observation} but now showing the mean significance when 1 year of data are analyzed. Although the trends remain the same independent of observation time, only the Arctic seems to be showing evidence of a tilted rotation axis throughout all thermal channels after one year of observation. For Antarctica, the result is wavelength dependent and suggest nine favourable channels. The difference between day and night measurements for the three equatorial regions are statistically none significant.}
        \label{Pic: PSD 1yr Observation}
\end{figure}

The Sahara Desert (a water-poor region), shows a strong signal (above 11$\sigma$) for the first six channels (20-25, 03.66-04.55~\textmu m) that are primarily used for surface, cloud and atmospheric temperature measurements, followed by five channels (29-32 \& 34, 08.40-12.27~\textmu m \& 13.49-13.79~\textmu m) that deliver an intermediate signal strength (above 6.5$\sigma$) and five channels (27-28 \& 33, 35-36, 06.54-07.48~\textmu m \& 13.19-14.39~\textmu m) with less signal strength (below 4.22$\sigma$). These bandwidths are primarily used for cirrus clouds, water vapour detection, and cloud top altitude detection. 

Besides from the above mentioned channels 20-23 that produce strong signals (11.83-12.15$\sigma$), we can deduce from the Indian Ocean (water rich region) power spectral densities that channels 27, 28 and 36 produce also strong signals (above 7$\sigma$) as well as that another four channels (30, 33-35) do show a peak at 1 cycle per year but with a low sigma reading (below 4.15$\sigma$), implying that in case of observations of a water dominated hemisphere at similar latitude bandwidths covering water absorption bands are less appropriate for detecting evidence of planetary obliquity via thermal emission monitoring. 

Since it is very unlikely to observe a rocky, temperate exoplanet around a sun-like star sufficiently frequently along its orbit over a time baseline of 15 years, we investigated the signal strength behaviour for an observation period of only one year. The same DFT analysis was performed separately on each year. For channels that did not show a peak at one cycle per year $\pm 5 \%$, the peak was retrieved from the signal at that orbital period. For each channel, the average of its signal strength throughout the 15 years and the corresponding standard deviation was calculated. The result of the analysis is shown in Figure \ref{Pic: PSD 1yr Observation}.

After one year of observations, the Arctic data set showed up to 3$\sigma$ signals for all channels except for channel 27 (06.54 - 06.90 \textmu m) which was the most variable channel in terms of radiance fluctuations. This is likely to be the result of undetectable cirrus clouds. 
Due to the relative large noise level and a very low reading (1.71$\sigma$) in the year 2010, it shows the largest errorbar out of the 16 channels. Only nine channels (24-28, 30, 33-36) lie above the 3$\sigma$ threshold for the Antarctic target location. Overall the channels show more noise than the matching channels of the Arctic data set, resulting in larger standard deviations. 

The three equatorial regions display the same behaviour for one year of observation as for 15 years of observation. Especially for channels 27 and 28 in the Sahara Desert and Indian Ocean data sets, as discussed above. None of the channels exceeded the 3$\sigma$ threshold for the locations in question considering only one year of observation. Their large standard deviations indicate the variability in spectral radiance due to a mixture of cloud coverage and change in seasonal insolation. 
For these latitudes however, the signal is probably dominated by clouds as they affect the measurements more due to a larger contrast between surface and top of cloud temperature. For the polar regions on the other hand, the signal is not dominated by clouds but by change of seasonal insolation. Indeed studies suggest that cloud coverage varies with distance from the equator where the cloudiest regions are the tropics and temperate zones and the subtropics and the polar regions have between 10-20\%  less cover \citep[e.g.,][]{Rossow1995, Rossow1999, Warren2007}. Furthermore, the results show that the difference between day and night measurements for the three equatorial locations is statistically insignificant in this type of analysis.

While the methodology to infer obliquity in the reflected light is well established \citep[e.g.,][]{Kawahara2016, Schwartz2016, Farr2018}, we cannot yet provide quantitative statements on the planet's obliquity based on our results.  
The underlying idea of our analysis is that by re-observing the planet several times along its orbit (and ideally over more than one orbital period) we can link the observed periodicity in the SED with the orbital period. 
Therefore we can identify whether a planet has seasons and hence a tilted spin axis. Additional and more quantitative conclusions require disk integrated data and additional modelling. We hope to address this in future research.


\section{Summary \& Conclusions}
\label{Sec: Conclusion}

We investigated the thermal emission and time variability of five single-surface-type target locations on Earth: 1) Amazon Rainforest, (2) Antarctica, (3) Arctic, (4) Indian Ocean and (5) Sahara Desert. 
We constructed Earth observation data sets containing calibrated spectral radiances from 16 discrete channels between 3.66-14.40~\textmu m over a time baseline of 15 years including separate data sets for day-time and night-time emission of all five locations. The data were collected between 2003 and 2017 by MODIS aboard NASA's Earth observing satellite Aqua which observed every target location twice per month, resulting in 360 measurements per location. From each data set, we derived spectral energy distributions and used Fourier analysis to look for evidence of planetary obliquity in the measured spectral radiance variations. By fitting black-body curves to the SEDs, we estimated the effective temperature for each target location. 
Since all our results are based on a time baseline of 15 consecutive years, we compare the time variability results to an observation time of one year in order to investigate whether it is still possible to detect evidence of Earth's tilted rotation axis.

\noindent Key results from the inferred SEDs include:
\begin{enumerate} 
\item As expected, Earth's thermal emission cannot be represented by a single SED. Hence, viewing geometry plays an important role when analyzing exoplanet thermal emission data. At Earth's peaking wavelength, the flux between the equatorial and polar regions varies on average by a factor of about two and three during the day and night, respectively.

\item Despite the coarse spectral resolution, the inferred SEDs showed differences in the strengths of absorption features in the atmosphere. Regions close to the equator and towards the north pole show more dominant O$_{3}$ and CO$_{2}$ absorption features than the south pole.

\item Given how seasonally variable the atmospheric structure and composition is in the Antarctica, its SED changes over half of an orbit significantly when the ozone feature at 9.65 \textmu m is compared between summer and winter. The lack of a clear ozone absorption feature during summer should be considered as it could lead to a false negative interpretation when searching for biosignatures in exoplanets that are viewed pole-on.

\item The derived effective temperatures averaged over a 15 year observation period are:
	\begin{itemize}
	\item[-] 278.01 K $\pm$ 0.31 K, 267.92 K $\pm$ 0.21 K and 260.52 $\pm$ 0.20 K for the three equatorial regions Sahara Desert, Indian Ocean and Amazon Rainforest during the day and 264.39 K $\pm$ 0.20 K, 266.82 K $\pm$ 0.20 K and 255.83 $\pm$ 0.17K during the night, respectively.
	\item[-] For the two poles we received 249.18 $\pm$ 0.16 and 236.75 $\pm$ 0.12 for the Arctic and 233.83 $\pm$ 0.06 and 219.00 $\pm$ 0.02 for Antarctica during summer and winter, respectively.
	\end{itemize}

\end{enumerate}

\noindent Key results from the time variability study over the 15-year observation period include:
\begin{enumerate}
\item Applying a DFT to the data sets showed that a periodicity of Earth’s thermal emission can be observed, providing a strong indication for a tilted spin axis, and therefore seasons.

\item Specifically, 92\% of the combined 128 thermal channels show a well defined peak at 1 cycle / year:
    \begin{enumerate}
    \item The polar regions displayed a signal strength of at least 10$\sigma$ for all channels. Therefore, pole-on view measurements are ideal for probing seasonal variations.
    \item The equatorial regions showed a signal strength of at least  7$\sigma$ for 72 out of 96 thermal channels, including the night measurements. 
    \end{enumerate}
\end{enumerate}

\noindent Key results from the time variability study over a one year observation period include:
\begin{enumerate}

\item For an observation time of one year ($\sim$24 data points), the strength of the periodicity signal decreases significantly. For the equatorial regions none of the investigated channels reached a 3$\sigma$-signal, and the day and night difference is found to be  statistically insignificant. 
For the polar regions, it is possible to observe signs of obliquity in the thermal emission but the significance depends on the spectral band considered.

\item Compared to the 15-year observation period, the trends of the data distribution are unchanged.

\item Comparing the nature of variability between the equatorial and polar regions, we find that the measurements are dominated by clouds in equatorial regions and by changes of seasonal insolation in polar regions.
\end{enumerate}

Our analyses emphasize the power of thermal emission data for the characterization of habitable terrestrial exoplanets, but they also caution us to be careful in the analysis and interpretation. 
Depending on viewing geometry and underlying dominant surface type the same planet may appear significantly different. It seems that seasons could be inferred from thermal emission data, but that will require a significant amount of observing time. Still, combining the information from all channels investigated here may allow to infer the existence of seasons with high confidence in case $\sim$20 data points are available roughly evenly spread over the planet's orbital period. 
The results obtained in this work should be considered as first steps towards disk-integrated mid-infrared observations of Earth’s spectra \citep[e.g.,][]{Hearty2009}, in order to explore characteristics that may appear in newly detected exoplanets in the future. The derived data sets could serve as a basis for such further investigations. We plan to perform a disk-integrated analysis in a follow up paper.
Furthermore, combining the optical and near-infrared channels of MODIS, that were not considered here, with the thermal emission data would allow us to investigate, for instance, the specific influence of various cloud types on certain channels in order to refine atmospheric models.

\acknowledgments
We thank an anonymous referee for the valuable comments. SPQ and RH thank the Swiss National Science Foundation (SNSF). Part of this work has been carried out within the framework of the National Centre for Competence in Research PlanetS supported by the SNFS. The authors also thank Tyler D. Robinson, Edward W. Schwieterman and Michael Schaepman for valuable comments. The MODIS/Aqua Calibrated Radiances 5-Min L1B Swath 1km (MYD021KM) data sets were acquired from the Level-1 and Atmosphere Archive \& Distribution System (LAADS) Distributed Active Archive Center (DAAC), located in the Goddard Space Flight Center in Greenbelt, Maryland \hyperlink{https://ladsweb.nascom.nasa.gov/}{https://ladsweb.nascom.nasa.gov/}.

\vspace{5mm}


\newpage
\appendix
\section{Additional Material}
\label{Appendix}

\begin{table}[htb!]
\centering
\caption{Resulting mean spectral radiances from the SED analyses in section \ref{SubSec: SEDs}. All values are stated in physical units of W m$^{-2}$\textmu m$^{-1}$ sr$^{-1}$.}
\label{Table: SED Values APPENDIX}
\small
\begin{tabular}{rlrlrlrlrlrlr}
\toprule
\toprule
 &  & \multicolumn{10}{c}{\textbf{Locations}} \\
 &  & \multicolumn{2}{c}{\textbf{Sahara Desert}} & \multicolumn{2}{c}{\textbf{Indian Ocean}} & \multicolumn{2}{c}{\textbf{Amazon Rainforest}}& \multicolumn{2}{c}{\textbf{Arctic}} & \multicolumn{2}{c}{\textbf{Antarctica}} \\
 &  & \multicolumn{1}{r}{Day} & \multicolumn{1}{l}{Night} & \multicolumn{1}{r}{Day} & \multicolumn{1}{l}{Night} & \multicolumn{1}{r}{Day} & \multicolumn{1}{l}{Night} & \multicolumn{1}{c}{Summer} & \multicolumn{1}{c}{Winter} & \multicolumn{1}{c}{Summer} & \multicolumn{1}{c}{Winter} \\
 \midrule
 & \textbf{20} &1.19&  0.25&  0.43&  0.30&  0.46&  0.21&    0.15&  0.04&  0.05&  0.01  \\
 & \textbf{21} &1.33&  0.36&  0.55&  0.44&  0.55&  0.32&    0.17&  0.05&  0.05&  0.01  \\
 & \textbf{22} &1.42&  0.35&  0.54&  0.43&  0.56&  0.31&    0.16&  0.05&  0.05&  0.01  \\
 & \textbf{23} &1.23&  0.40&  0.53&  0.46&  0.54&  0.33&    0.15&  0.06&  0.05&  0.02  \\
 & \textbf{24} &0.16&  0.11&  0.13&  0.12&  0.12&  0.10&    0.08&  0.04&  0.06&  0.02  \\
 & \textbf{25} &0.62&  0.37&  0.40&  0.38&  0.36&  0.30&    0.17&  0.09&  0.07&  0.04  \\
 & \textbf{27} &1.46&  1.39&  1.58&  1.52&  1.17&  1.15&    1.06&  0.90&  0.84&  0.62  \\
 & \textbf{28} &3.07&  2.85&  3.09&  2.99&  2.35&  2.39&    2.06&  1.59&  1.29&  0.82  \\
 & \textbf{29} &9.17&  6.21&  7.00&  6.79&  6.16&  5.47&    3.99&  2.76&  2.14&  1.24  \\
 & \textbf{30} &6.66&  4.90&  5.41&  5.30&  5.02&  4.58&    3.34&  2.23&  2.69&  1.38  \\
 & \textbf{31} &10.29&  7.23&  7.59&  7.40& 6.70&  6.12&    4.89&  3.61&  2.91& 1.88  \\
 & \textbf{32} &9.49&  6.91&  7.08&  6.91&  6.20&  5.74&    4.78&  3.63&  2.95&  1.99  \\
 & \textbf{33} &5.47&  4.79&  4.83&  4.76&  4.30&  4.16&    3.81&  3.10&  2.79&  2.05  \\
 & \textbf{34} &4.17&  3.92&  3.97&  3.94&  3.63&  3.55&    3.33&  2.72&  2.65&  2.00  \\
 & \textbf{35} &3.56&  3.34&  3.39&  3.37&  3.16&  3.11&    3.05&  2.44&  2.60&  1.92  \\
\multirow{-17}{*}{\rotatebox[origin=c]{90}{\textbf{Channels}}} & \textbf{36} &3.00&  2.93&  2.95&  2.94&  2.86& 2.85&  3.12&  2.56&  2.91&  2.30 \\
\bottomrule
\end{tabular}
\end{table}

\bibliography{mybib_MSC.bib}

\begin{thebibliography}{}
\expandafter\ifx\csname natexlab\endcsname\relax\def\natexlab#1{#1}\fi
\providecommand{\url}[1]{\href{#1}{#1}}

\bibitem[{{Alei} {et~al.}(2020){Alei}, {Claudi}, {Bignamini}, \&
  {Molinaro}}]{alei2020}
{Alei}, E., {Claudi}, R., {Bignamini}, A., \& {Molinaro}, M. 2020, arXiv
  e-prints, arXiv:2002.01834

\bibitem[{{Anglada-Escud{\'e}} {et~al.}(2016){Anglada-Escud{\'e}}, {Amado},
  {Barnes}, {Berdi{\~n}as}, {Butler}, {Coleman}, {de La Cueva}, {Dreizler},
  {Endl}, {Giesers}, {Jeffers}, {Jenkins}, {Jones}, {Kiraga}, {K{\"u}rster},
  {L{\'o}pez-Gonz{\'a}lez}, {Marvin}, {Morales}, {Morin}, {Nelson}, {Ortiz},
  {Ofir}, {Paardekooper}, {Reiners}, {Rodr{\'{\i}}guez},
  {Rodr{\'{\i}}guez-L{\'o}pez}, {Sarmiento}, {Strachan}, {Tsapras}, {Tuomi}, \&
  {Zechmeister}}]{Anglada2016}
{Anglada-Escud{\'e}}, G., {Amado}, P.~J., {Barnes}, J., {et~al.} 2016, \nat,
  536, 437

\bibitem[{Arney {et~al.}(2016)Arney, Domagal-Goldman, Meadows, Wolf,
  Schwieterman, Charnay, Claire, Hébrard, \& Trainer}]{Arney2016}
Arney, G., Domagal-Goldman, S.~D., Meadows, V.~S., {et~al.} 2016, Astrobiology,
  16, 873–899.
\newblock \url{http://dx.doi.org/10.1089/ast.2015.1422}

\bibitem[{Batalha(2014)}]{Batalha2014}
Batalha, N.~M. 2014, Proceedings of the National Academy of Sciences, 111,
  12647.
\newblock \url{https://www.pnas.org/content/111/35/12647}

\bibitem[{Burke {et~al.}(2015)Burke, Christiansen, Mullally, Seader, Huber,
  Rowe, Coughlin, Thompson, Catanzarite, Clarke, Morton, Caldwell, Bryson,
  Haas, Batalha, Jenkins, Tenenbaum, Twicken, Li, Quintana, Barclay, Henze,
  Borucki, Howell, \& Still}]{Burke2015}
Burke, C., Christiansen, J., Mullally, F., {et~al.} 2015, Astrophysical
  Journal, 809, doi:10.1088/0004-637X/809/1/8

\bibitem[{Butchart(2014)}]{Butchart2014}
Butchart, N. 2014, Reviews of Geophysics, 52, 157.
\newblock
  \url{https://agupubs.onlinelibrary.wiley.com/doi/abs/10.1002/2013RG000448}

\bibitem[{Catling {et~al.}(2018)Catling, Krissansen-Totton, Kiang, Crisp,
  Robinson, DasSarma, Rushby, Del~Genio, Bains, \&
  Domagal-Goldman}]{Catling2018}
Catling, D.~C., Krissansen-Totton, J., Kiang, N.~Y., {et~al.} 2018,
  Astrobiology, 18, 709, pMID: 29676932.
\newblock \url{https://doi.org/10.1089/ast.2017.1737}

\bibitem[{Christensen \& Pearl(1997)}]{Christensen1997}
Christensen, P.~R., \& Pearl, J.~C. 1997, Journal of Geophysical Research:
  Planets, 102, 10875.
\newblock
  \url{https://agupubs.onlinelibrary.wiley.com/doi/abs/10.1029/97JE00637}

\bibitem[{{Cowan} {et~al.}(2012){Cowan}, {Voigt}, \& {Abbot}}]{Cowan2012}
{Cowan}, N.~B., {Voigt}, A., \& {Abbot}, D.~S. 2012, \apj, 757, 80

\bibitem[{Cowan {et~al.}(2009)Cowan, Agol, Meadows, Robinson, Livengood,
  Deming, Lisse, A'Hearn, Wellnitz, Seager, Charbonneau, \& the
  EPOXI~Team}]{Cowan2009}
Cowan, N.~B., Agol, E., Meadows, V.~S., {et~al.} 2009, The Astrophysical
  Journal, 700, 915.
\newblock \url{http://stacks.iop.org/0004-637X/700/i=2/a=915}

\bibitem[{Dittmann {et~al.}(2017)Dittmann, Irwin, Charbonneau, Bonfils,
  Astudillo-Defru, Haywood, Berta-Thompson, Newton, Rodriguez, Winters, Tan,
  Almenara, Bouchy, Delfosse, Forveille, Lovis, Murgas, Pepe, Santos, Udry,
  W{\"u}nsche, Esquerdo, Latham, \& Dressing}]{Dittmann2017}
Dittmann, J.~A., Irwin, J.~M., Charbonneau, D., {et~al.} 2017, Nature, 544, 333
  EP .
\newblock \url{https://doi.org/10.1038/nature22055}

\bibitem[{Farr {et~al.}(2018)Farr, Farr, Cowan, Haggard, \&
  Robinson}]{Farr2018}
Farr, B., Farr, W.~M., Cowan, N.~B., Haggard, H.~M., \& Robinson, T. 2018, The
  Astronomical Journal, 156, 146

\bibitem[{{Ford} {et~al.}(2001){Ford}, {Seager}, \& {Turner}}]{Ford2001}
{Ford}, E.~B., {Seager}, S., \& {Turner}, E.~L. 2001, \nat, 412, 885

\bibitem[{Fujii {et~al.}(2011)Fujii, Kawahara, Suto, Fukuda, Nakajima,
  Livengood, \& Turner}]{Fujii2011}
Fujii, Y., Kawahara, H., Suto, Y., {et~al.} 2011, The Astrophysical Journal,
  738, 184.
\newblock \url{https://doi.org/10.1088%2F0004-637x%2F738%2F2%2F184}

\bibitem[{Gaudi {et~al.}(2020)Gaudi, Seager, Mennesson, Kiessling, Warfield,
  Cahoy, Clarke, Domagal-Goldman, Feinberg, Guyon, Kasdin, Mawet, Plavchan,
  Robinson, Rogers, Scowen, Somerville, Stapelfeldt, Stark, Stern, Turnbull,
  Amini, Kuan, Martin, Morgan, Redding, Stahl, Webb, Alvarez-Salazar, Arnold,
  Arya, Balasubramanian, Baysinger, Bell, Below, Benson, Blais, Booth,
  Bourgeois, Bradford, Brewer, Brooks, Cady, Caldwell, Calvet, Carr, Chan,
  Cormarkovic, Coste, Cox, Danner, Davis, Dewell, Dorsett, Dunn, East,
  Effinger, Eng, Freebury, Garcia, Gaskin, Greene, Hennessy, Hilgemann, Hood,
  Holota, Howe, Huang, Hull, Hunt, Hurd, Johnson, Kissil, Knight, Kolenz,
  Kraus, Krist, Li, Lisman, Mandic, Mann, Marchen, Marrese-Reading, McCready,
  McGown, Missun, Miyaguchi, Moore, Nemati, Nikzad, Nissen, Novicki, Perrine,
  Pineda, Polanco, Putnam, Qureshi, Richards, Riggs, Rodgers, Rud, Saini,
  Scalisi, Scharf, Schulz, Serabyn, Sigrist, Sikkia, Singleton, Shaklan, Smith,
  Southerd, Stahl, Steeves, Sturges, Sullivan, Tang, Taras, Tesch, Therrell,
  Tseng, Valente, Buren, Villalvazo, Warwick, Webb, Westerhoff, Wofford, Wu,
  Woo, Wood, Ziemer, Arney, Anderson, Maíz-Apellániz, Bartlett, Belikov,
  Bendek, Cenko, Douglas, Dulz, Evans, Faramaz, Feng, Ferguson, Follette, Ford,
  García, Geha, Gelino, Götberg, Hildebrandt, Hu, Jahnke, Kennedy, Kreidberg,
  Isella, Lopez, Marchis, Macri, Marley, Matzko, Mazoyer, McCandliss, Meshkat,
  Mordasini, Morris, Nielsen, Newman, Petigura, Postman, Reines, Roberge,
  Roederer, Ruane, Schwieterman, Sirbu, Spalding, Teplitz, Tumlinson, Turner,
  Werk, Wofford, Wyatt, Young, \& Zellem}]{Gaudi2020}
Gaudi, B.~S., Seager, S., Mennesson, B., {et~al.} 2020, The Habitable Exoplanet
  Observatory (HabEx) Mission Concept Study Final Report,  arXiv e-print,
  arXiv:2001.06683

\bibitem[{Gilbert {et~al.}(2020)Gilbert, Barclay, Schlieder, Quintana, Hord,
  Kostov, Lopez, Rowe, Hoffman, Walkowicz, Silverstein, Rodriguez, Vanderburg,
  Suissa, Airapetian, Clement, Raymond, Mann, Kruse, Lissauer, Colón, kumar
  Kopparapu, Kreidberg, Zieba, Collins, Quinn, Howell, Ziegler, Vrijmoet,
  Adams, Arney, Boyd, Brande, Burke, Cacciapuoti, Chance, Christiansen, Covone,
  Daylan, Dineen, Dressing, Essack, Fauchez, Galgano, Howe, Kaltenegger, Kane,
  Lam, Lee, Lewis, Logsdon, Mandell, Monsue, Mullally, Mullally, Paudel,
  Pidhorodetska, Plavchan, Reyes, Rinehart, Rojas-Ayala, Smith, Stassun,
  Tenenbaum, Vega, Villanueva, Wolf, Youngblood, Ricker, Vanderspek, Latham,
  Seager, Winn, Jenkins, Bakos, Briceño, Ciardi, Cloutier, Conti, Couperus,
  Sora, Eisner, Everett, Gan, Hartman, Henry, Isopi, Jao, Jensen, Law, Mallia,
  Matson, Shappee, Wood, \& Winters}]{Gilbert2020}
Gilbert, E.~A., Barclay, T., Schlieder, J.~E., {et~al.} 2020, The First
  Habitable Zone Earth-sized Planet from TESS. I: Validation of the TOI-700
  System, , , arXiv:2001.00952

\bibitem[{Gillon {et~al.}(2017)Gillon, Triaud, Demory, Jehin, Agol, Deck,
  Lederer, de~Wit, Burdanov, Ingalls, Bolmont, Leconte, Raymond, Selsis,
  Turbet, Barkaoui, Burgasser, Burleigh, Carey, Chaushev, Copperwheat, Delrez,
  Fernandes, Holdsworth, Kotze, Van~Grootel, Almleaky, Benkhaldoun, Magain, \&
  Queloz}]{Gillon2017}
Gillon, M., Triaud, A. H. M.~J., Demory, B.-O., {et~al.} 2017, Nature, 542,
  doi:10.1038/nature21360;.
\newblock \url{http://dx.doi.org/10.1038/nature21360}

\bibitem[{{G{\'o}mez-Leal} {et~al.}(2016){G{\'o}mez-Leal}, {Codron}, \&
  {Selsis}}]{Gomez2016}
{G{\'o}mez-Leal}, I., {Codron}, F., \& {Selsis}, F. 2016, \icarus, 269, 98

\bibitem[{{G{\'o}mez-Leal} {et~al.}(2012){G{\'o}mez-Leal}, {Pall{\'e}}, \&
  {Selsis}}]{Gomez2012}
{G{\'o}mez-Leal}, I., {Pall{\'e}}, E., \& {Selsis}, F. 2012, \apj, 752, 28

\bibitem[{Hanel {et~al.}(1972)Hanel, Conrath, Kunde, Prabhakara, Revah,
  Salomonson, \& Wolford}]{Hanel1972}
Hanel, R.~A., Conrath, B.~J., Kunde, V.~G., {et~al.} 1972, Journal of
  Geophysical Research, 77, 2629.
\newblock
  \url{https://agupubs.onlinelibrary.wiley.com/doi/abs/10.1029/JC077i015p02629}

\bibitem[{Hearty {et~al.}(2009)Hearty, Song, Kim, \& Tinetti}]{Hearty2009}
Hearty, T., Song, I., Kim, S., \& Tinetti, G. 2009, The Astrophysical Journal,
  693, 1763.
\newblock \url{http://stacks.iop.org/0004-637X/693/i=2/a=1763}

\bibitem[{Hudson \& Brandt(2005)}]{Hudson2005}
Hudson, S.~R., \& Brandt, R.~E. 2005, Journal of Climate, 18, 1673.
\newblock \url{https://doi.org/10.1175/JCLI3360.1}

\bibitem[{Hurley {et~al.}(2014)Hurley, Irwin, Adriani, Moriconi, Oliva,
  Capaccioni, Smith, Filacchione, Tosi, \& Thomas}]{HURLEY2014}
Hurley, J., Irwin, P., Adriani, A., {et~al.} 2014, Planetary and Space Science,
  90, 37 .
\newblock
  \url{http://www.sciencedirect.com/science/article/pii/S0032063313001530}

\bibitem[{{International Earth rotation and Reference systems
  Service}(2014)}]{IERS}
{International Earth rotation and Reference systems Service}. 2014,
  International Earth rotation and Reference systems,  online source, [Online;
  accessed 21-December-2018].
\newblock \url{https://www.iers.org/IERS/EN/}

\bibitem[{Jiang {et~al.}(2018)Jiang, Zhai, Herman, Zhai, Hu, Su, Natraj, Li,
  Xu, \& Yung}]{Jiang2018}
Jiang, J.~H., Zhai, A.~J., Herman, J., {et~al.} 2018, The Astronomical Journal,
  156, 26.
\newblock \url{https://doi.org/10.3847%2F1538-3881%2Faac6e2}

\bibitem[{Jurgenson {et~al.}(2016)Jurgenson, Fischer, McCracken, Sawyer,
  Szymkowiak, Davis, Muller, \& Santoro}]{Jurgenson_2016}
Jurgenson, C., Fischer, D., McCracken, T., {et~al.} 2016, in Ground-based and
  Airborne Instrumentation for Astronomy VI, ed. C.~J. Evans, L.~Simard, \&
  H.~Takami, Vol. 9908, International Society for Optics and Photonics (SPIE),
  2051 -- 2070.
\newblock \url{https://doi.org/10.1117/12.2233002}

\bibitem[{Kaltenegger {et~al.}(2007)Kaltenegger, Traub, \&
  Jucks}]{Kaltenegger_2007}
Kaltenegger, L., Traub, W.~A., \& Jucks, K.~W. 2007, The Astrophysical Journal,
  658, 598–616.
\newblock \url{http://dx.doi.org/10.1086/510996}

\bibitem[{Kasting \& Catling(2003)}]{Kasting_2003}
Kasting, J.~F., \& Catling, D. 2003, Annual Review of Astronomy and
  Astrophysics, 41, 429.
\newblock \url{https://doi.org/10.1146/annurev.astro.41.071601.170049}

\bibitem[{Kawahara(2016)}]{Kawahara2016}
Kawahara, H. 2016, The Astrophysical Journal, 822, 112.
\newblock \url{http://dx.doi.org/10.3847/0004-637X/822/2/112}

\bibitem[{{King} {et~al.}(2013){King}, {Platnick}, {Menzel}, {Ackerman}, \&
  {Hubanks}}]{King2013}
{King}, M.~D., {Platnick}, S., {Menzel}, W.~P., {Ackerman}, S.~A., \&
  {Hubanks}, P.~A. 2013, IEEE Transactions on Geoscience and Remote Sensing,
  51, 3826

\bibitem[{Kondratyev(1999)}]{Kondratyev1999}
Kondratyev, K.~Y. 1999, Climatic Effects of Aerosols and Clouds (Heidelberg:
  Springer)

\bibitem[{{Krissansen-Totton} {et~al.}(2018){Krissansen-Totton}, {Garland},
  {Irwin}, \& {Catling}}]{totton2018}
{Krissansen-Totton}, J., {Garland}, R., {Irwin}, P., \& {Catling}, D.~C. 2018,
  \aj, 156, 114

\bibitem[{Lin {et~al.}(2019)Lin, Wolfe, Zhang, Tilton, Dellomo, \&
  Tan}]{Lin2019}
Lin, G.~G., Wolfe, R.~E., Zhang, P., {et~al.} 2019, in Earth Observing Systems
  XXIV, ed. J.~J. Butler, X.~J. Xiong, \& X.~Gu, Vol. 11127, International
  Society for Optics and Photonics (SPIE), 219 -- 230.
\newblock \url{https://doi.org/10.1117/12.2529447}

\bibitem[{Livengood {et~al.}(2011)Livengood, Deming, A'Hearn, Charbonneau,
  Hewagama, Lisse, McFadden, Meadows, Robinson, Seager, \&
  Wellnitz}]{Livengood2011}
Livengood, T.~A., Deming, L.~D., A'Hearn, M.~F., {et~al.} 2011, Astrobiology,
  11, 907.
\newblock \url{https://doi.org/10.1089/ast.2011.0614}

\bibitem[{Mayor \& Queloz(1995)}]{Mayor1995}
Mayor, M., \& Queloz, D. 1995, Nature, 378, 355 EP .
\newblock \url{https://doi.org/10.1038/378355a0}

\bibitem[{Meadows(2008)}]{Meadows_2008}
Meadows, V. 2008, Planetary Environmental Signatures for Habitability and Life
  (Springer Praxis Books. Springer, Berlin, Heidelberg)

\bibitem[{{MODIS User Guide}(2017)}]{MODIS_Userguide}
{MODIS User Guide}. 2017, MODIS Level 1B Product User’s Guide.
\newblock
  \url{https://mcst.gsfc.nasa.gov/sites/default/files/file_attachments/M1054E_PUG_2017_0901_V6.2.2_Terra_V6.2.1_Aqua.pdf}

\bibitem[{Molina \& Rowland(1974)}]{Molina1974}
Molina, M.~J., \& Rowland, F.~S. 1974, Nature, 249, 810 EP .
\newblock \url{https://doi.org/10.1038/249810a0}

\bibitem[{{Montet} {et~al.}(2015){Montet}, {Morton}, {Foreman-Mackey},
  {Johnson}, {Hogg}, {Bowler}, {Latham}, {Bieryla}, \& {Mann}}]{Montet2015}
{Montet}, B.~T., {Morton}, T.~D., {Foreman-Mackey}, D., {et~al.} 2015, \apj,
  809, 25

\bibitem[{{Morley} {et~al.}(2017){Morley}, {Kreidberg}, {Rustamkulov},
  {Robinson}, \& {Fortney}}]{morley2017}
{Morley}, C.~V., {Kreidberg}, L., {Rustamkulov}, Z., {Robinson}, T., \&
  {Fortney}, J.~J. 2017, \apj, 850, 121

\bibitem[{Olson {et~al.}(2018)Olson, Schwieterman, Reinhard, Ridgwell, Kane,
  Meadows, \& Lyons}]{Olson2018}
Olson, S.~L., Schwieterman, E.~W., Reinhard, C.~T., {et~al.} 2018, The
  Astrophysical Journal, 858, L14.
\newblock \url{https://doi.org/10.3847%2F2041-8213%2Faac171}

\bibitem[{Pall{\'e} {et~al.}(2009)Pall{\'e}, Osorio, Barrena,
  Monta{\~n}{\'e}s-Rodr{\'\i}guez, \& Mart{\'\i}n}]{Palle_2009}
Pall{\'e}, E., Osorio, M. R.~Z., Barrena, R., Monta{\~n}{\'e}s-Rodr{\'\i}guez,
  P., \& Mart{\'\i}n, E.~L. 2009, Nature, 459, 814.
\newblock \url{https://doi.org/10.1038/nature08050}

\bibitem[{Parkinson(2003)}]{Aqua_Brochure}
Parkinson, C.~L. 2003, Aqua - Brochure.
\newblock \url{https://www.nasa.gov/pdf/151986main_Aqua_brochure.pdf}

\bibitem[{Petitcolin \& Vermote(2002)}]{Petitcolin2002}
Petitcolin, F., \& Vermote, E. 2002, Remote Sensing of Environment, 83, 112 ,
  the Moderate Resolution Imaging Spectroradiometer (MODIS): a new generation
  of Land Surface Monitoring.
\newblock
  \url{http://www.sciencedirect.com/science/article/pii/S0034425702000949}

\bibitem[{{Quanz} {et~al.}(2015){Quanz}, {Crossfield}, {Meyer}, {Schmalzl}, \&
  {Held}}]{quanz2015}
{Quanz}, S.~P., {Crossfield}, I., {Meyer}, M.~R., {Schmalzl}, E., \& {Held}, J.
  2015, International Journal of Astrobiology, 14, 279

\bibitem[{{Quanz} {et~al.}(2018){Quanz}, {Kammerer}, {Defr{\`e}re}, {Absil},
  {Glauser}, \& {Kitzmann}}]{Quanz2018}
{Quanz}, S.~P., {Kammerer}, J., {Defr{\`e}re}, D., {et~al.} 2018, in Society of
  Photo-Optical Instrumentation Engineers (SPIE) Conference Series, Vol. 10701,
  \procspie, 107011I

\bibitem[{{Quirrenbach} {et~al.}(2014){Quirrenbach}, {Amado}, {Caballero},
  {Mundt}, {Reiners}, {Ribas}, {Seifert}, {Abril}, {Aceituno},
  {Alonso-Floriano}, {Ammler-von Eiff}, {Antona Jim{\'e}nez}, {Anwand
  -Heerwart}, {Azzaro}, {Bauer}, {Barrado}, {Becerril}, {B{\'e}jar},
  {Ben{\'\i}tez}, {Berdi{\~n}as}, {C{\'a}rdenas}, {Casal}, {Claret},
  {Colom{\'e}}, {Cort{\'e}s-Contreras}, {Czesla}, {Doellinger}, {Dreizler},
  {Feiz}, {Fern{\'a}ndez}, {Galad{\'\i}}, {G{\'a}lvez-Ortiz},
  {Garc{\'\i}a-Piquer}, {Garc{\'\i}a-Vargas}, {Garrido}, {Gesa}, {G{\'o}mez
  Galera}, {Gonz{\'a}lez {\'A}lvarez}, {Gonz{\'a}lez Hern{\'a}ndez},
  {Gr{\"o}zinger}, {Gu{\`a}rdia}, {Guenther}, {de Guindos},
  {Guti{\'e}rrez-Soto}, {Hagen}, {Hatzes}, {Hauschildt}, {Helmling}, {Henning},
  {Hermann}, {Hern{\'a}ndez Casta{\~n}o}, {Herrero}, {Hidalgo}, {Holgado},
  {Huber}, {Huber}, {Jeffers}, {Joergens}, {de Juan}, {Kehr}, {Klein},
  {K{\"u}rster}, {Lamert}, {Lalitha}, {Laun}, {Lemke}, {Lenzen}, {L{\'o}pez del
  Fresno}, {L{\'o}pez Mart{\'\i}}, {L{\'o}pez-Santiago}, {Mall}, {Mandel},
  {Mart{\'\i}n}, {Mart{\'\i}n-Ruiz}, {Mart{\'\i}nez-Rodr{\'\i}guez}, {Marvin},
  {Mathar}, {Mirabet}, {Montes}, {Morales Mu{\~n}oz}, {Moya}, {Naranjo},
  {Ofir}, {Oreiro}, {Pall{\'e}}, {Panduro}, {Passegger}, {P{\'e}rez-Calpena},
  {P{\'e}rez Medialdea}, {Perger}, {Pluto}, {Ram{\'o}n}, {Rebolo}, {Redondo},
  {Reffert}, {Reinhardt}, {Rhode}, {Rix}, {Rodler}, {Rodr{\'\i}guez},
  {Rodr{\'\i}guez-L{\'o}pez}, {Rodr{\'\i}guez-P{\'e}rez}, {Rohloff}, {Rosich},
  {S{\'a}nchez-Blanco}, {S{\'a}nchez Carrasco}, {Sanz-Forcada}, {Sarmiento},
  {Sch{\"a}fer}, {Schiller}, {Schmidt}, {Schmitt}, {Solano}, {Stahl}, {Storz},
  {St{\"u}rmer}, {Su{\'a}rez}, {Ulbrich}, {Veredas}, {Wagner}, {Winkler},
  {Zapatero Osorio}, {Zechmeister}, {Abell{\'a}n de Paco},
  {Anglada-Escud{\'e}}, {del Burgo}, {Klutsch}, {Lizon}, {L{\'o}pez-Morales},
  {Morales}, {Perryman}, {Tulloch}, \& {Xu}}]{Quirrenbach_2014}
{Quirrenbach}, A., {Amado}, P.~J., {Caballero}, J.~A., {et~al.} 2014, Society
  of Photo-Optical Instrumentation Engineers (SPIE) Conference Series, Vol.
  9147, {CARMENES instrument overview}, 91471F

\bibitem[{Rauer {et~al.}(2014)Rauer, Catala, Aerts, Appourchaux, Benz,
  Brandeker, Christensen-Dalsgaard, Deleuil, Gizon, Goupil, \&
  et~al.}]{Rauer_2014}
Rauer, H., Catala, C., Aerts, C., {et~al.} 2014, Experimental Astronomy, 38,
  249–330.
\newblock \url{http://dx.doi.org/10.1007/s10686-014-9383-4}

\bibitem[{Reinhard {et~al.}(2017)Reinhard, Olson, Schwieterman, \&
  Lyons}]{Reinhard_2017}
Reinhard, C.~T., Olson, S.~L., Schwieterman, E.~W., \& Lyons, T.~W. 2017,
  Astrobiology, 17, 287, pMID: 28418704.
\newblock \url{https://doi.org/10.1089/ast.2016.1598}

\bibitem[{{Ricker} {et~al.}(2015){Ricker}, {Winn}, {Vanderspek}, {Latham},
  {Bakos}, {Bean}, {Berta-Thompson}, {Brown}, {Buchhave}, {Butler}, {Butler},
  {Chaplin}, {Charbonneau}, {Christensen-Dalsgaard}, {Clampin}, {Deming},
  {Doty}, {De Lee}, {Dressing}, {Dunham}, {Endl}, {Fressin}, {Ge}, {Henning},
  {Holman}, {Howard}, {Ida}, {Jenkins}, {Jernigan}, {Johnson}, {Kaltenegger},
  {Kawai}, {Kjeldsen}, {Laughlin}, {Levine}, {Lin}, {Lissauer}, {MacQueen},
  {Marcy}, {McCullough}, {Morton}, {Narita}, {Paegert}, {Palle}, {Pepe},
  {Pepper}, {Quirrenbach}, {Rinehart}, {Sasselov}, {Sato}, {Seager},
  {Sozzetti}, {Stassun}, {Sullivan}, {Szentgyorgyi}, {Torres}, {Udry}, \&
  {Villasenor}}]{TESS_2015}
{Ricker}, G.~R., {Winn}, J.~N., {Vanderspek}, R., {et~al.} 2015, Journal of
  Astronomical Telescopes, Instruments, and Systems, 1, 014003

\bibitem[{Robinson(2011)}]{Robinson2011}
Robinson, T.~D. 2011, The Astrophysical Journal, 741, 51.
\newblock \url{https://doi.org/10.1088%2F0004-637x%2F741%2F1%2F51}

\bibitem[{Robinson {et~al.}(2014)Robinson, Ennico, Meadows, Sparks, Bussey,
  Schwieterman, \& Breiner}]{Robinson2014}
Robinson, T.~D., Ennico, K., Meadows, V.~S., {et~al.} 2014, The Astrophysical
  Journal, 787, 171.
\newblock \url{http://stacks.iop.org/0004-637X/787/i=2/a=171}

\bibitem[{{Robinson} \& {Reinhard}(2018)}]{Robinson2018}
{Robinson}, T.~D., \& {Reinhard}, C.~T. 2018, arXiv e-prints, arXiv:1804.04138

\bibitem[{Rossow \& Schiffer(1999)}]{Rossow1999}
Rossow, W.~B., \& Schiffer, R.~A. 1999, Bull. Amer. Meteorol. Soc., 80, 2261

\bibitem[{Rossow \& Zhang(1995)}]{Rossow1995}
Rossow, W.~B., \& Zhang, Y.-C. 1995, J. Geophys. Res., 100, 1167

\bibitem[{Rugheimer \& Kaltenegger(2018)}]{Rugheimer_2018}
Rugheimer, S., \& Kaltenegger, L. 2018, The Astrophysical Journal, 854, 19.
\newblock \url{https://doi.org/10.3847%2F1538-4357%2Faaa47a}

\bibitem[{Sagan {et~al.}(1993)Sagan, Thompson, Carlson, Gurnett, \&
  Hord}]{Sagan1993}
Sagan, C., Thompson, W.~R., Carlson, R., Gurnett, D., \& Hord, C. 1993, Nature,
  365, 715 EP .
\newblock \url{https://doi.org/10.1038/365715a0}

\bibitem[{Schwartz {et~al.}(2016)Schwartz, Sekowski, Haggard, Pallé, \&
  Cowan}]{Schwartz2016}
Schwartz, J.~C., Sekowski, C., Haggard, H.~M., Pallé, E., \& Cowan, N.~B.
  2016, Monthly Notices of the Royal Astronomical Society, 457, 926–938.
\newblock \url{http://dx.doi.org/10.1093/mnras/stw068}

\bibitem[{Schwieterman {et~al.}(2018)Schwieterman, Kiang, Parenteau, Harman,
  DasSarma, Fisher, Arney, Hartnett, Reinhard, Olson, \&
  et~al.}]{Schwieterman2018}
Schwieterman, E.~W., Kiang, N.~Y., Parenteau, M.~N., {et~al.} 2018,
  Astrobiology, 18, 663–708.
\newblock \url{http://dx.doi.org/10.1089/ast.2017.1729}

\bibitem[{{Snellen} {et~al.}(2015){Snellen}, {de Kok}, {Birkby}, {Brandl},
  {Brogi}, {Keller}, {Kenworthy}, {Schwarz}, \& {Stuik}}]{snellen2015}
{Snellen}, I., {de Kok}, R., {Birkby}, J.~L., {et~al.} 2015, \aap, 576, A59

\bibitem[{{The LUVOIR Team}(2019)}]{LUVOIR_2019}
{The LUVOIR Team}. 2019, arXiv e-prints, arXiv:1912.06219

\bibitem[{Tremblay {et~al.}(2020)Tremblay, Line, Stevenson, Kataria, Zellem,
  Fortney, \& Morley}]{Tremblay2020}
Tremblay, L., Line, M.~R., Stevenson, K., {et~al.} 2020, The Astronomical
  Journal, 159, 117.
\newblock \url{https://iopscience.iop.org/article/10.3847/1538-3881/ab64dd}

\bibitem[{Turnbull {et~al.}(2006)Turnbull, Traub, Jucks, Woolf, Meyer, Gorlova,
  Skrutskie, \& Wilson}]{Turnbull2006}
Turnbull, M.~C., Traub, W.~A., Jucks, K.~W., {et~al.} 2006, The Astrophysical
  Journal, 644, 551.
\newblock \url{http://stacks.iop.org/0004-637X/644/i=1/a=551}

\bibitem[{Warren {et~al.}(2007)Warren, Eastman, \& Hahn}]{Warren2007}
Warren, S.~G., Eastman, R.~M., \& Hahn, C.~J. 2007, Journal of Climate, 20,
  717.
\newblock \url{https://doi.org/10.1175/JCLI4031.1}

\bibitem[{Yang {et~al.}(2018)Yang, Marshak, V{\'a}rnai, \&
  Knyazikhin}]{Yang2018}
Yang, W., Marshak, A., V{\'a}rnai, T., \& Knyazikhin, Y. 2018, in Remote
  Sensing

\bibitem[{Young {et~al.}(2018)Young, Naylor, Brake, Fisher, Seneta, Kunst,
  Pepe, \& Sosnowska}]{Young_2018}
Young, J., Naylor, T., Brake, M., {et~al.} 2018, in Software and
  Cyberinfrastructure for Astronomy V, ed. J.~C. Guzman \& J.~Ibsen, Vol.
  10707, International Society for Optics and Photonics (SPIE), 732 -- 741.
\newblock \url{https://doi.org/10.1117/12.2312830}

\end{thebibliography}



\end{document}